\documentclass[preprint,12pt]{elsarticle}

\usepackage{amssymb}
\usepackage{amsmath}
\usepackage{amsthm}

\usepackage{graphicx}% Include figure files
\usepackage{dcolumn}% Align table columns on decimal point
\usepackage{bm}% bold math
\usepackage{tikz}
\usepackage{physics}
\usepackage{qcircuit}
\usepackage{multirow}
\usepackage{intcalc}
\usepackage{hyperref}
\usepackage{algorithm}
\usepackage{algpseudocode}
\usepackage{ulem}
\usepackage{xcolor}
\newtheorem*{theorem}{Property}

\newcommand{\dnqv}{$\text{D}n\text{Q}v$ }

\journal{Computers and fluids}

\begin{document}

\begin{frontmatter}



\title{Quantum collision circuit, quantum invariants and quantum phase estimation procedure for fluid dynamic lattice gas automata}


\author[M2P2,LIS]{Niccolò Fonio} %% Author name
\ead{niccolo.fonio@lis-lab.fr}

\affiliation[M2P2]{organization={M2P2, Aix-Marseille University, Central Marseille},%Department and Organization
            addressline={M2P2 UMR 7340, 38 rue Joliot-Curie}, 
            city={Marseille},
            postcode={13013}, 
            state={},
            country={France}}

\affiliation[LIS]{organization={LIS, Aix-Marseille university},%Department and Organization
            addressline={LIS UMR 7020, Campus de Luminy, 163 avenue de Luminy}, 
            city={Marseille},
            postcode={13288}, 
            state={},
            country={France}}

\author[M2P2]{Pierre Sagaut}
\author[LIS]{Giuseppe Di Molfetta}

%% Abstract
\begin{abstract}
Lattice Gas Cellular Automata (LGCA) is a classical numerical method widely known and applied to simulate several physical phenomena. In this paper, we study the translation of LGCA on quantum computers (QC) using computational basis encoding (CBE), developing methods for different purposes. In particular, we clarify and discuss some fundamental limitations and advantages in using CBE and quantum walk as streaming procedure. Using quantum walks affect the possible encoding of classical states in quantum orthogonal states, feature linked to the unitarity of collision and to the possibility of getting a quantum advantage. Then, we give efficient procedures for optimizing collisional quantum circuits, based on the classical features of the model. This is applied specifically to fluid dynamic LGCA. Alongside, a new collision circuit for a 1-dimensional model is proposed. We address the important point of invariants in LGCA providing a method for finding how many invariants appear in their QC formulation. Quantum invariants outnumber the classical expectations, proving the necessity of further research. Lastly, we prove the validity of a method for retrieving any quantity of interest based on quantum phase estimation (QPE).
\end{abstract}

%\includegraphics[scale=0.3]{graphic_abstract.eps}

%\item  Quantum collision circuit for fluid dynamic simulation
%\item Protocol for retrieving classical information avoiding reinitialization
%\item Analysis of conserved quantities in Quantum Lattice Gas
%\item Limitations of sublinear encoding of the space in Quantum Lattice Gas

%% Keywords
\begin{keyword}
Cellular Automata \sep Lattice gas \sep Quantum Computing \sep Quantum Circuits \sep Fluid-dynamic


\end{keyword}

\end{frontmatter}



\section{\label{sec:intro} Introduction}

A \textit{Lattice Gas Cellular Automata} (LGCA), addressed as \dnqv model, is a gas of particles propagating in a discretized space of $n$ dimensions where the components of the system (particles) can exhibit $v$ discrete velocities~\cite{wolf2004lattice}. Each \dnqv model consists of a lattice and a discrete evolution rule. The lattice gas is encoded with a bit string, i.e. \textit{cell}, for each lattice point. In every cell each bit represents the presence of a particle with the corresponding velocity, as in Fig.\ref{fig:D1Q3_lattice_noevolution}.
\begin{figure}[ht]
    \centering
    \scalebox{0.85}{
    \begin{tikzpicture}
        \def\l{0.5}
        
        \newcommand{\drawlattice}{
        \foreach \j in {0}{
        \foreach \i in {0,1,2,3,4,5}
                {
                \draw[fill=white] (\i, \j) circle (3pt);
                }
            }
        }
        
        \newcommand{\drawarrows}[3]{
        \foreach \ang in {#3}{
        \draw[-stealth, black, thick] 
        ({#1 + 0.1 * cos(\ang)}, #2) -- ({#1 + \l*cos(\ang)},#2);}}
        \newcommand{\drawrest}[2]{
        \draw[fill=black] (#1, #2) circle (3pt);
        }
        
        \drawlattice
        \drawarrows{1}{0}{0,180}
        \drawrest{2}{0}
        \drawarrows{3}{0}{0}
        \drawarrows{4}{0}{180}
        \drawrest{5}{0}
        \drawrest{4}{0}
        \filldraw[black] (6, 0) node[anchor=center]{$\dots$};
        \drawarrows{7}{0}{0,180}
        \drawrest{7}{0}
        
        \filldraw[black] (0, -0.75) node[anchor=center]{$[000]$};
        \filldraw[black] (1, -0.75) node[anchor=center]{$[101]$};
        \filldraw[black] (2, -0.75) node[anchor=center]{$[010]$};
        \filldraw[black] (3, -0.75) node[anchor=center]{$[001]$};
        \filldraw[black] (4, -0.75) node[anchor=center]{$[110]$};
        \filldraw[black] (5, -0.75) node[anchor=center]{$[010]$};
        \filldraw[black] (6, -0.75) node[anchor=center]{$\dots$};
        \filldraw[black] (7, -0.75) node[anchor=center]{$[111]$};
        \filldraw[black] (0, 0.75) node[anchor=center]{$0$};
        \filldraw[black] (1, 0.75) node[anchor=center]{$1$};
        \filldraw[black] (2, 0.75) node[anchor=center]{$2$};
        \filldraw[black] (3, 0.75) node[anchor=center]{$3$};
        \filldraw[black] (4, 0.75) node[anchor=center]{$4$};
        \filldraw[black] (5, 0.75) node[anchor=center]{$5$};
        \filldraw[black] (6, 0.75) node[anchor=center]{$\dots$};
        \filldraw[black] (7, 0.75) node[anchor=center]{$N$};
        
        \filldraw[black] (-2, 0.75) node[anchor=center]{$x$};
        \filldraw[black] (-2, 0) node[anchor=center]{lattice gas};
        \filldraw[black] (-2, -0.75) node[anchor=center]{encoding};
        
    \end{tikzpicture}
    }
    \caption{Example of a 1D lattice gas with $N$ cells and 3 velocities $[-1,0,+1]$. The presence of a rest particle is a black dot. The presence of a moving particle is an arrow.}
    \label{fig:D1Q3_lattice_noevolution}
\end{figure}
The evolution rule is made of a collision step when particles scatter in each cell enforcing some conservation laws, and a streaming step when particles move to neighboring cells according to their velocity. The LGCA models we are going to consider, introduced in Sec.\ref{sec:classical models}, are a 1D model, namely D1Q3, and a 2D model, namely D2Q6. The latter has been studied by Friesch, Hasslacher and Pomeau~\cite{frisch1986lattice}, and thus it is addressed as FHP. The evolution of a LGCA is capable of simulating various physical non-linear phenomena ~\cite{doolen2019lattice} and, most interestingly for us, they can be used for computational fluid dynamics (CFD). In particular FHP has been the first LGCA model capable of retrieving Navier-Stokes-like equations. For the opportunities of \dnqv models for CFD and given the possible advantages that quantum computing (QC) is showing directly in this field ~\cite{bharadwaj2020quantum, succi2023quantum}, this paper focuses on the quantum computing formulation of LGCA, named herinafter QLGCA.

It is possible to look at QLGCA from different perspectives. First, they can be seen as a subclass of quantum cellular autoamata (QCA) \cite{farrelly2020review}, considering \textit{qubits} instead of bits. This has been the contribution of some seminal works \cite{meyer1997quantum, meyer1997quantum2}, that provide no advantage in terms of computational efficiency, but show the possibility of simulating PDEs using quantum systems. Another perspective is to look at QLGCA and compare them with the corresponding classical numerical methods. In this case, we are interested in getting an advantage in terms of computational resources (number of qubits and operations), and we aim to find the best representation of the classical system for a quantum algorithm. In this sense, much attention has been given to \textit{quantum Lattice Boltzmann Methods} (QLBM).

LBM \cite{kruger2016lattice,mohamad2011lattice} is a family of classical lattice-based  numerical methods that were born from LGCA, and solved some issues for hydrodynamic simulations (e.g. preservation of Galilean invariance), becoming one of the most used CFD methods. Instead of microscopic particles, in LBM we stream mesoscopic probability distribution functions, solving the lattice Boltzmann equation under specific assumptions and simplifications.
A significant difference with LGCA is that, in most cases, the collision term of LBM is nonlinear. The first QLBM was developed by Yepez ~\cite{yepez2001quantum, yepez2001quantum2, yepez2002efficient, yepez2002quantum} and showed that it was possible to replicate LBM on a quantum computer and to solve PDEs. Since then, different approaches have been proposed and are in rapid expansion.
One approach is to linearize the non-linear collision operator of LBM adopting Carleman linearization~\cite{itani2024quantum, itani2022analysis}. Another one gets an exponential advantage in space complexity, requiring measurement and reinitialization at each time step~\cite{budinski2021quantum, budinski2021quantum2}. These first alternatives are probabilistic since they rely on a linear combination of unitaries~\cite{childs2012hamiltonian, low2019hamiltonian}. Alternative encodings and applications can be found, each with different advantages and drawbacks ~\cite{todorova2020quantum, schalkers2024importance}. 
The principle of each QLBM proposed is to interpret the quantum amplitudes as the classical probability distribution functions. Despite being one of the most promising foreseen ways for QCFD, the non-unitarity of the process and the necessity of measurements and reinitialization hinder a real advantage at the current state of the art. Thus, it is worth looking at features and methods for QLGCA, as we are going to discuss in this paper.

In particular, LGCA can exhibit 2 advantages over LBM for QC implementation: (P1) the collision consists of a non-deterministic or deterministic correspondence of input/output states, and (P2) the collision is the same in each cell. The property (P2) allows to leverage quantum parallelism, detailed in Sec.\ref{subsec:encodings}, for a computational advantage. The property (P1) can be advantageous in QC assuming the computational basis encoding (CBE) that we introduce in Sec.\ref{subsec:encodings}. This encoding of classical states into quantum states allows to carry out deterministic collisions as unitary operations, and non-deterministic collisions as unitary operations followed by measurements. Unitary operations are the fundamental operations used in QC, and performing unitary collision and streaming with a small number of qubits is a crucial aspect of the quantum advantage. 
Usually, this is not possible for the non-unitarity of the collisions, as in \cite{budinski2021quantum, budinski2021quantum2, succi2023quantum}. 
QLGCA with CBE, on the other hand, shows a true correspondence between linearity of the model and unitarity of the collision. This is promising for the development of a \textit{multi-timestep} algorithm, thus without need for measurement and reinitialization as computational parts of the algorithm. However, it presents other challenges.

In fact, the CBE we use has already shown some limitations, as outlined in \cite{schalkers2024importance}. It was proved that using CBE for a D1Q3 model does not allow for streaming and collision representing the space with a logarithmic number of qubits. We show in Sec.\ref{sec:methods} that this limitation goes beyond the CBE used in \cite{schalkers2024importance}, including any encoding of the classical states in a set of orthonormal quantum states, also called \textit{orthogonal states encoding} (OSE). Considering that these limitations hinder the achievement of a quantum advantage at the current state, our results prove that this is not strictly linked to the CBE, but to the coexistence of CBE and the streaming procedure using quantum walks. Thus, future research may still prove an advantage considering CBE or OSE to which all the methods in this paper can be applied.

Regarding the property (P1), we said that we can translate the classical collision process into a unitary operator. The decomposition of this operator into quantum gates, which are the basic operations in QC, is necessary for executing the algorithm on real devices and can be expensive \cite{nielsen2010quantum}. We propose and prove the validity of different algorithmic and quantum algorithmic methods for carrying out the collision of D1Q3 in Sec.\ref{sec:d1q3} and some collisions of FHP in Sec.\ref{sec:fhp} with an optimal decomposition in quantum gates. In particular for FHP we use an operator whose default decomposition using Qiskit\cite{javadi2024quantum} needs an order of $10^4$ operations, while our methods find an overall quantum circuit that uses just 291 operations, giving an improvement of $\approx 100$ times. 

In addition to computational costs, another crucial aspect when transitioning from the classical model to the quantum model regards the conservation laws. LGCA are based on preserving specific quantities, called \textit{invariants}, such as mass and momentum. In QC, quantities are considered to be Hermitian operators, called \textit{observables}. An observable is conserved, being called \textit{quantum invariant}, if it commutes with the collision operator \cite{love2019quantum}. In classical and quantum LGCA there can be some \textit{spurious invariants} \cite{wolf2004lattice, zanetti1989hydrodynamics, zanetti1991counting, bernardin1992global}, which are anomalous conserved quantities that can affect the behavior of the system with additional coupled conservation equations. In Sec.\ref{sec:methods} we develop a method for counting the number of quantum invariants given a collision operator, and we discover that spurious quantum invariants are numerous compared to the classical case. This finding questions the "simple" translation from classical quantities to Hermitian operators, i.e. quantum observables, and opens perspectives never tackled in previous works, highlighting the need of further research on this topic.

The existence of quantum observables and the unexpected number of quantum invariant show the wide possibilities of QLGCA, going to the core of interpreting information in LGCA. Thus, the last aspect we address regards the retrieving of information. Classically, information is accessible at any moment, making it feasible to know the state of the lattice at each time step. In QC retrieving information is possible through measurements, that directly affect the quantum state. Procedures of (re)initialization and measurements were proposed for QLBM algorithms \cite{budinski2021quantum, budinski2021quantum2, zamora2025efficient}. These procedures, despite being promising, rely on efficient protocols that have yet to be discovered and are necessary for assessing quantum advantage. 
In our paper, we introduce a novel approach for retrieving quantities of interest, such as mass and momentum, using a \textit{quantum phase estimation} (QPE) algorithm \cite{kitaev1995quantum,nielsen2010quantum}. This algorithm allows information to be accessed probabilistically during the computation without necessarily measure the quantum state of the cell. This does not solve the problem of measurement and reinitialization for an advantageous encoding, but is a general method applicable to any quantity of interest for any algorithm using CBE. 

To summarize, the contributions of this paper are fourfold:
\begin{itemize}
    \item expand the fundamental limitations of adopting CBE and quantum walk streaming protocol;
    \item we develop an optimal collision circuit for FHP and a novel collision circuit for D1Q3;
    \item we develop a new method for counting conserved quantities in quantum LGCA, applying it to D1Q3 and FHP and showing unexpected results;
    \item propose novel QPE protocols for retrieving important physical quantities.
\end{itemize}
The paper is structured as follows: in Sec.\ref{sec:classical models} we introduce extensively the classical models; in Sec.\ref{sec:methods} we introduce the CBE and the encodings of the space, discussing the advantages and the methods we develop; in Sec.\ref{sec:d1q3} we apply the methods to D1Q3 and in Sec.\ref{sec:fhp} to FHP.

\section{\label{sec:classical models}Classical models}
\subsection*{D1Q3}
The first LGCA we consider is D1Q3. D1Q3 is a 1-dimensional lattice gas with 3 velocities: a right-moving particle of mass 1, a left-moving particle of mass 1, and a rest particle of mass 2. Applying an exclusion principle, such that only one particle per velocity per site is allowed, each site is represented with a bit-string $[n_2n_1n_0]$. Conventionally, $n_2,n_1,n_0$ represent the presence of a particle with velocity, respectively, $-1,0,1$. This system can simulate diffusion, and it is the easiest model to perform mass- and momentum-conserving collision, ensuring non-linearities. The collision allowed (and its inverse) consists of splitting a rest particle into two opposite-moving particles (the opposite is merging two opposite-moving particles into a rest particle). This can be represented as $\qty[101] \leftrightarrow \qty[010]$. Fig.\ref{fig:D1Q3_lattice} provides an example of a 1-time step evolution of this gas.

\begin{figure}[ht]
    \centering
    \scalebox{0.85}{
    \begin{tikzpicture}
        \def\l{0.5}
        
        \newcommand{\drawlattice}{
        \foreach \j in {0,1.5,3}{
        \foreach \i in {0,1,2,3,4,5}
                {
                \draw[fill=white] (\i, \j) circle (3pt);
                }
            }
        }
        
        \newcommand{\drawarrows}[3]{
        \foreach \ang in {#3}{
        \draw[-stealth, black, thick] 
        ({#1 + 0.1 * cos(\ang)}, #2) -- ({#1 + \l*cos(\ang)},#2);}}
        \newcommand{\drawrest}[2]{
        \draw[fill=black] (#1, #2) circle (3pt);
        }
        
        \drawlattice
        \drawarrows{1}{3}{0,180}
        \drawrest{2}{3}
        \drawarrows{3}{3}{0}
        \drawarrows{4}{3}{180}
        \drawrest{5}{3}
        \drawrest{4}{3}
        \filldraw[black] (0, 2.5) node[anchor=center]{$[000]$};
        \filldraw[black] (1, 2.5) node[anchor=center]{$[101]$};
        \filldraw[black] (2, 2.5) node[anchor=center]{$[010]$};
        \filldraw[black] (3, 2.5) node[anchor=center]{$[001]$};
        \filldraw[black] (4, 2.5) node[anchor=center]{$[110]$};
        \filldraw[black] (5, 2.5) node[anchor=center]{$[010]$};
        \draw[gray, thick] (-3.5,2) -- (5.5,2);
        \drawrest{1}{1.5}
        \drawarrows{2}{1.5}{0,180}
        \drawarrows{3}{1.5}{0}
        \drawarrows{4}{1.5}{180}
        \drawarrows{5}{1.5}{0,180}
        \drawrest{4}{1.5}
        \filldraw[black] (0, 1) node[anchor=center]{$[000]$};
        \filldraw[black] (1, 1) node[anchor=center]{$[010]$};
        \filldraw[black] (2, 1) node[anchor=center]{$[101]$};
        \filldraw[black] (3, 1) node[anchor=center]{$[001]$};
        \filldraw[black] (4, 1) node[anchor=center]{$[110]$};
        \filldraw[black] (5, 1) node[anchor=center]{$[101]$};
        \draw[gray, thick] (-3.5,0.5) -- (5.5,0.5);
        \drawrest{1}{0}
        \drawarrows{0}{0}{0}
        \drawarrows{1}{0}{180}
        \drawarrows{3}{0}{0,180}
        \drawarrows{4}{0}{0,180}
        \drawarrows{5}{0}{180}
        \drawrest{4}{0}
        \filldraw[black] (0, -0.5) node[anchor=center]{$[001]$};
        \filldraw[black] (1, -0.5) node[anchor=center]{$[110]$};
        \filldraw[black] (2, -0.5) node[anchor=center]{$[000]$};
        \filldraw[black] (3, -0.5) node[anchor=center]{$[101]$};
        \filldraw[black] (4, -0.5) node[anchor=center]{$[111]$};
        \filldraw[black] (5, -0.5) node[anchor=center]{$[100]$};

        \filldraw[black] (-2, 4) node[anchor=center]{$x$};
        \filldraw[black] (0, 4) node[anchor=center]{$0$};
        \filldraw[black] (1, 4) node[anchor=center]{$1$};
        \filldraw[black] (2, 4) node[anchor=center]{$2$};
        \filldraw[black] (3, 4) node[anchor=center]{$3$};
        \filldraw[black] (4, 4) node[anchor=center]{$4$};
        \filldraw[black] (5, 4) node[anchor=center]{$5$};
        
        \filldraw[black] (-2, 2.75) node[anchor=center]{Before collision};
        \filldraw[black] (-2, 1.25) node[anchor=center]{After collision};
        \filldraw[black] (-2, -0.25) node[anchor=center]{After streaming};
    \end{tikzpicture}
    }
    \caption{D1Q3 evolution example. In this case, each cell is represented with 3 bits $[n_2n_1n_0]$. Before the collision in the cell at $x=1,2,5$ there are collisional configurations, so they evolve according to the chosen collision. All the other cells are not affected by the collision. The streaming takes place according to respective velocities}
    \label{fig:D1Q3_lattice}
\end{figure}

Classically, we can write pseudocode as in Alg.\ref{alg:cap} for the collision step. Here we can anticipate that the quantum analogous of a code or psudocode is a quantum gate-based circuit, interpreting operations as series of quantum gates.
\begin{algorithm} [ht]
\caption{Collision D1Q3 algorithm}\label{alg:cap}
\begin{algorithmic}
\For {cell $\in$ lattice}
\State $\text{m} \gets \text{get mass(cell)}$
\Comment{Retrieving of quantities}
\State $\text{p} \gets \text{get momentum(cell)}$
\If{cell $=[010]$ or cell $=[101]$}
\State $\text{cell} \gets \text{new state}$ \Comment{Collision process}
\EndIf
\EndFor
\end{algorithmic}
\end{algorithm}
Quantities such as mass $m(x)=\sum_i n_i(x)$ and momentum $p(x)=n_0(x) - n_1(x)$ are conserved. Averaging over neighbors we get mass density $\rho(x)$ and momentum density $\Vec{u}(x)$. The behavior of these quantities in the continuum limit follows differential equations, which depend on the conservation laws applied.

\subsection*{FHP}
The second LGCA we consider, which is more interesting for computational fluid dynamics (CFD) simulations, was proven by Frisch, Hasslacher, and Pomeau (FHP)\cite{frisch1986lattice} and Wolfram~\cite{wolfram1986cellular} to converge in the continuous limit to the Navier-Stokes equations. It is a 2-dimensional LGCA on a triangular lattice, showing a collision that changes the configurations of the cell, followed by a streaming step. An example of the evolution step is drawn in Fig. \ref{fig:fhp_evol}.

\begin{figure}[ht]
    \centering
    \scalebox{0.85}{
        \begin{tikzpicture}
        \def\r{0.8}
        \def\D{0}
        \def\R{0.6}

        \newcommand{\drawarrows}[4]{
        \foreach \ang in {#3}{
        \draw[-stealth, black, thick] 
        ({#4 + #1*\r + 0.5*\r * \intcalcMod{#2}{2}+0.13*cos(\ang)},{0.866*#2*\r+0.13*sin(\ang)}) -- ({#4 + #1*\r + 0.5*\r * \intcalcMod{#2}{2}+\R*cos(\ang)},{0.866*#2*\r+\R*sin(\ang)});}}

        \newcommand{\drawlattice}[1]{
        \foreach \j in {0,1,2,3,4,5}{
        \foreach \i in {0,1,2,3,4}{
        \draw[fill=white] (#1 + \i*\r + 0.5*\r * \intcalcMod{\j}{2},0.866*\j*\r ) circle (3pt);}}}
        
        \drawlattice{0}
        \drawarrows{0}{0}{0,60,120,180,300,360}{0}
        \drawarrows{4}{4}{60,180,300}{0}
        \drawarrows{2}{3}{60,240}{0}
        \drawarrows{1}{4}{120,180,240}{0}
        \drawarrows{3}{1}{0,120,180,300}{0}
        
        \drawlattice{5}
        \drawarrows{0}{0}{0,60,120,180,300,360}{5}
        \drawarrows{4}{4}{0,120,240}{5}
        \drawarrows{2}{3}{0,180}{5}
        \drawarrows{1}{4}{120,180,240}{5}
        \drawarrows{3}{1}{60,180,240,0}{5}

        \drawlattice{10}
        \drawarrows{1}{0}{0}{10}
        \drawarrows{3}{0}{240}{10}
        \drawarrows{4}{0}{180}{10}
        
        \drawarrows{0}{1}{60}{10}
        \drawarrows{2}{1}{180}{10}
        \drawarrows{4}{1}{0,120}{10}

        \drawarrows{4}{2}{60}{10}

        \drawarrows{0}{3}{240}{10}
        \drawarrows{1}{3}{180}{10}
        \drawarrows{3}{3}{0,240}{10}

        \drawarrows{0}{4}{0,180}{10}

        \drawarrows{0}{5}{120,300}{10}
        \drawarrows{3}{5}{120}{10}

        \draw[gray, thick] (4.25,-1) -- (4.25,4.75);
        \draw[gray, thick] (9.25,-1) -- (9.25,4.75);
        \filldraw[black] (1.75, 4.5) node[anchor=center]{\Large Before collision};
        \filldraw[black] (6.75, 4.5) node[anchor=center]{\Large After collision};
        \filldraw[black] (11.75, 4.5) node[anchor=center]{\Large After streaming};
        \end{tikzpicture}
    }
    
    \caption{Evolution step for Frisch, Hasslacher, and Pomeau model (FHP). Before collision is the starting state of the lattice. In the center, we see that the particles in cells with the collisional states of Table.\ref{tab:collisions2} are rearranged. In the last part we see that particles have been streamed to neighboring cells applying periodic boundary conditions}
    \label{fig:fhp_evol}
\end{figure}
The set of velocities $\qty{\Vec{c}_i}$ is defined as follows, and each $c_i$ is linked to $n_i$ and ordered as $[n_5 n_4 n_3 n_2 n_1 n_0]$

\begin{equation}
    \vec{c}_i = \qty(cos \qty(\frac{\pi}{3}i),sin \qty(\frac{\pi}{3}i) ) \text{  for } i=0,1,\dots,5    
\end{equation} 
Among different elastic collisions, we focus on 0-momentum collisions represented in Table\ref{tab:collisions2}. These are the collisions that involve cells with a configuration that has momentum $p=0$, and they are represented in Table \ref{tab:collisions2}. A rest particle can be added for stability of the numerical simulations and for adding new collisions, causing a faster thermalization to the equilibrium of the solution.

\begin{table}[t]
\centering
\begin{tabular}{m{0.2\textwidth} | m{0.2\textwidth} | m{0.2\textwidth}}
\centering $B_2$&
\centering $B_3$&
$B_4$\\
\centering
\scalebox{0.40}{
\begin{tikzpicture}
    \draw[-stealth, black, thick] (0.13,0) -- (1,0);
    \draw[-stealth, white, thick] (0.065,0.11258) -- (0.5,0.866);
    \draw[-stealth, white, thick] (-0.065,0.11258) -- (-0.5,0.866);
    \draw[-stealth, black, thick] (-0.13,0) -- (-1,0);
    \draw[-stealth, white, thick] (-0.065,-0.11258) -- (-0.5,-0.866);
    \draw[-stealth, white, thick] (0.065,-0.11258) -- (0.5,-0.866);
    \draw[black] (0,0) circle (4pt);

    \def\dx{4}
    \def\dy{0}
    \draw[-stealth, white, thick] (0.13+\dx,0+\dy) -- (1+\dx,0+\dy);
    \draw[-stealth, black, thick] (0.065+\dx,0.11258+\dy) -- (0.5+\dx,0.866+\dy);
    \draw[-stealth, white, thick] (-0.065+\dx,0.11258+\dy) -- (-0.5+\dx,0.866+\dy);
    \draw[-stealth, white, thick] (-0.13+\dx,0+\dy) -- (-1+\dx,0+\dy);
    \draw[-stealth, black, thick] (-0.065+\dx,-0.11258+\dy) -- (-0.5+\dx,-0.866+\dy);
    \draw[-stealth, white, thick] (0.065+\dx,-0.11258+\dy) -- (0.5+\dx,-0.866+\dy);
    \draw[black] (0+\dx,0+\dy) circle (4pt);

    \def\dx{2}
    \def\dy{0.866*4}
    \draw[-stealth, white, thick] (0.13+\dx,0+\dy) -- (1+\dx,0+\dy);
    \draw[-stealth, white, thick] (0.065+\dx,0.11258+\dy) -- (0.5+\dx,0.866+\dy);
    \draw[-stealth, black, thick] (-0.065+\dx,0.11258+\dy) -- (-0.5+\dx,0.866+\dy);
    \draw[-stealth, white, thick] (-0.13+\dx,0+\dy) -- (-1+\dx,0+\dy);
    \draw[-stealth, white, thick] (-0.065+\dx,-0.11258+\dy) -- (-0.5+\dx,-0.866+\dy);
    \draw[-stealth, black, thick] (0.065+\dx,-0.11258+\dy) -- (0.5+\dx,-0.866+\dy);
    \draw[black] (0+\dx,0+\dy) circle (4pt);

    \newcommand{\drawdoubarrow}[4]{
        \draw[-stealth, black, thick] (#1,#2) -- ({#1+#3*cos(#4)},{#2+sin(#4)*#3});
        \draw[-stealth, black, thick] ({#1+#3*cos(#4)+0.3},{#2+sin(#4)*#3}) -- (#1+0.3,#2);
    }

    \drawdoubarrow{0.65}{1.25}{1}{60}
    \drawdoubarrow{3.25}{1.25}{1}{120}

    \draw[-stealth, black, thick] (1.5,0.1) -- (2.5,0.1);
    \draw[-stealth, black, thick] (2.5,-0.1) -- (1.5,-0.1);

\end{tikzpicture}} & 
\centering
\scalebox{0.40}{
\begin{tikzpicture}
    \draw[-stealth, black, thick] (0.13,0) -- (1,0);
    %\draw[-stealth, black, thick] (0.065,0.11258) -- (0.5,0.866);
    \draw[-stealth, black, thick] (-0.065,0.11258) -- (-0.5,0.866);
    %\draw[-stealth, black, thick] (-0.13,0) -- (-1,0);
    \draw[-stealth, black, thick] (-0.065,-0.11258) -- (-0.5,-0.866);
    %\draw[-stealth, black, thick] (0.065,-0.11258) -- (0.5,-0.866);
    \draw[black] (0,0) circle (4pt);
    %\draw[fill=black] (-0.13,0) -- (0.13,0) arc(-0.13:180:0.13) --cycle;

    \draw[-stealth, black, thick] (1.5,0.1) -- (2.5,0.1);
    \draw[-stealth, black, thick] (2.5,-0.1) -- (1.5,-0.1);

    %\draw[-stealth, black, thick] (0.13,0) -- (1,0);
    \draw[-stealth, black, thick] (4.065,0.11258) -- (4.5,0.866);
    %\draw[-stealth, black, thick] (-0.065,0.11258) -- (-0.5,0.866);
    \draw[-stealth, black, thick] (3.87,0) -- (3,0);
    %\draw[-stealth, black, thick] (-0.065,-0.11258) -- (-0.5,-0.866);
    \draw[-stealth, black, thick] (4.065,-0.11258) -- (4.5,-0.866);
    \draw[black] (4,0) circle (4pt);
    %\draw[fill=black] (4-0.13,0) -- (4+0.13,0) arc(-0.13:180:0.13) --cycle;

    \draw[-stealth, white, thick] (2,-0.2) -- (2, -2.5);
    
\end{tikzpicture}} &
\centering
\scalebox{0.40}{
\begin{tikzpicture}
    \draw[-stealth, black, thick] (0.13,0) -- (1,0);
    \draw[-stealth, black, thick] (0.065,0.11258) -- (0.5,0.866);
    \draw[-stealth, white, thick] (-0.065,0.11258) -- (-0.5,0.866);
    \draw[-stealth, black, thick] (-0.13,0) -- (-1,0);
    \draw[-stealth, black, thick] (-0.065,-0.11258) -- (-0.5,-0.866);
    \draw[-stealth, white, thick] (0.065,-0.11258) -- (0.5,-0.866);
    \draw[black] (0,0) circle (4pt);

    \def\dx{4}
    \def\dy{0}
    \draw[-stealth, white, thick] (0.13+\dx,0+\dy) -- (1+\dx,0+\dy);
    \draw[-stealth, black, thick] (0.065+\dx,0.11258+\dy) -- (0.5+\dx,0.866+\dy);
    \draw[-stealth, black, thick] (-0.065+\dx,0.11258+\dy) -- (-0.5+\dx,0.866+\dy);
    \draw[-stealth, white, thick] (-0.13+\dx,0+\dy) -- (-1+\dx,0+\dy);
    \draw[-stealth, black, thick] (-0.065+\dx,-0.11258+\dy) -- (-0.5+\dx,-0.866+\dy);
    \draw[-stealth, black, thick] (0.065+\dx,-0.11258+\dy) -- (0.5+\dx,-0.866+\dy);
    \draw[black] (0+\dx,0+\dy) circle (4pt);

    \def\dx{2}
    \def\dy{0.866*4}
    \draw[-stealth, black, thick] (0.13+\dx,0+\dy) -- (1+\dx,0+\dy);
    \draw[-stealth, white, thick] (0.065+\dx,0.11258+\dy) -- (0.5+\dx,0.866+\dy);
    \draw[-stealth, black, thick] (-0.065+\dx,0.11258+\dy) -- (-0.5+\dx,0.866+\dy);
    \draw[-stealth, black, thick] (-0.13+\dx,0+\dy) -- (-1+\dx,0+\dy);
    \draw[-stealth, white, thick] (-0.065+\dx,-0.11258+\dy) -- (-0.5+\dx,-0.866+\dy);
    \draw[-stealth, black, thick] (0.065+\dx,-0.11258+\dy) -- (0.5+\dx,-0.866+\dy);
    \draw[black] (0+\dx,0+\dy) circle (4pt);

    \newcommand{\drawdoubarrow}[4]{
        \draw[-stealth, black, thick] (#1,#2) -- ({#1+#3*cos(#4)},{#2+sin(#4)*#3});
        \draw[-stealth, black, thick] ({#1+#3*cos(#4)+0.3},{#2+sin(#4)*#3}) -- (#1+0.3,#2);
    }

    \drawdoubarrow{0.65}{1.25}{1}{60}
    \drawdoubarrow{3.25}{1.25}{1}{120}

    \draw[-stealth, black, thick] (1.5,0.1) -- (2.5,0.1);
    \draw[-stealth, black, thick] (2.5,-0.1) -- (1.5,-0.1);

\end{tikzpicture}
}
\end{tabular}

\caption{\label{tab:collisions2}%
These are the 0-momentum collisions of the model FHP model. We can notice that $B_2$ and $B_4$ are probabilistic rotations of 120° or 240°, taking place with equal probability, while $B_3$ is a rotation of 180°. $B_2$ and $B_4$ collisional states are invariant under $B_3$, and vice versa
}

\end{table}

If any cell in the lattice is found to be in a collisional state represented in Table \ref{tab:collisions2}, then the scattering process takes place and particles are rearranged.  We can see these collisions as rotations. Rotations can also have invariant states. Trivially the full and empty cells are invariant respect to any rotation. We notice, as reported in Table \ref{tab:0asymmetric}, that the states involved in 2- and 4-body collisions, called $B_2$ and $B_4$, are invariant respect to a rotation of 180°, that is going to be the 3-body collision called $B_3$. These invariances are going to be used for the quantum implementation of the algorithm. Specific features used for finding the optimized quantum circuit are reported in Sec.\ref{subsec:fhp coll circ}.

Likewise the case of D1Q3, we show a quantum gate-based implementation for the collision step involving 0-momentum collisions. The quantities that are conserved classically are, as before, mass $m(x) = \sum_{i=0}^5 n_i(x)$ and momentum $\Vec{p}(x) = \sum_{i=0}^5 \Vec{c}_i n_i(x)$. These quantities are averaged over neighbouring cells and their evolution is mapped to the continuous limit with a Chapman-Enskog expansion. This retrieves Euler's equations at the first order and Navier-Stokes equations at the second order~\cite{wolf2004lattice,frisch1986lattice}. We define the same quantities as quantum observables, verifying that they are conserved and we show how we can count the total amount of quantum invariants given a specific collision.

\section{\label{sec:methods} Encodings and methods}

A classical \dnqv model needs $Nv$ bits to represent the state of the lattice with $N$ grid points. These are $Nv$ values of 0s or 1s, as in Fig.\ref{fig:D1Q3_lattice_noevolution}. The fundamental element in quantum computing, analogous to the classical bit, is the qubit. A qubit is a vector in a Hilbert space $\ket{\psi}\in\mathcal{H}^2$ that can be written as $\ket{\psi}=\alpha \ket{0} + \beta \ket{1}$, where $\ket{0}$ and $\ket{1}$ compose the \textit{computational basis}, and $\alpha,\beta \in \mathbb{C}$ and $|\alpha|^2 + |\beta|^2 = 1$. A qubit represents a quantum states where the physical system is \textit{at the same time} in $\ket{0}$ and $\ket{1}$. If we carry out a measurement, the system collapses in $\ket{0}$ with probability $|\alpha|^2$, or in $\ket{1}$ with probability $|\beta|^2$. Considering column vector notation $\ket{0}=(1,0)^T$ and $\ket{1}=(0,1)^T$, being $\ket{\psi}=(\alpha,\beta)^T$. Multiqubit states are represented with the cross product of the single qubits, usually implicit, and can use binary notation (e.g. a three-qubit state can be $\ket{0}\otimes \ket{1}\otimes \ket{0} = \ket{010} = \ket{2}$). To change the state of a qubit we need to apply \textit{unitary} operations $\hat{U}$, called \textit{gates}. These can be single-qubit gates, thus representable as $2\cross2$ unitary matrices, or $v-$qubit gates, representable as $2^v\cross2^v$ unitary matrices.
In this section, we present possible encoding of LGCA in quantum states and methods for three purposes: optimal implementation of collision, calculation of quantum invariants, and retrieving quantities of interest.

\subsection{\label{subsec:encodings}Encodings}

In this paper we use the Computational basis encoding (CBE). We encode the cell populated by particles with $v$ different velocities in a \dnqv model with $v$ qubits. We consider the quantum states $\ket{0}$ and $\ket{1}$ for the absence and presence of a particle with respective velocity. 
\begin{align}
    \text{Classical encoding}  & \longrightarrow \text{Quantum register} \notag \\ 
    \qty[n_v n_{v-1} \dots n_0] & \longrightarrow \ket{n_v n_{v-1} \dots n_0} \label{eq:cell encoding}
\end{align}
An encoding with $v$ qubits per cell is used in different works for QLBM \cite{yepez2001quantum, yepez2001quantum2, yepez2002efficient} and QLGA \cite{zamora2025efficient}, and our methods can be applied to different encodings of the space. In particular, we can consider a \textit{linear encoding} and a \textit{sublinear encoding} of the space. 

The linear encoding of the space utilizes $Nv$ qubits. With this encoding, the state $\ket{\Psi(t)}$ of the entire lattice at time $t$  is as follows
\begin{equation}
    \ket{\Psi(t)} = \bigotimes_{x=0}^N\ket{\psi(x,t)} = \bigotimes_{x=0}^N\ket{n_v(x,t) \dots n_0(x,t)}
\end{equation}
where $\ket{\psi(x,t)}$ is the state of the cell in $x$ at time $t$. This encoding does not allow for any advantage of interest for quantum simulations of classical CFD schemes. Nevertheless, this system corresponds to a Watrous quantum cellular automata \cite{farrelly2020review, watrous1995one}. Thus, the methods proposed here are of interest for quantum simulations of QCA. 

The sublinear encoding, on the other hand, leverages superposition, entanglement, and quantum parallelism to seek an advantage. If we have $n=\log_2N$ qubits, we are capable of representing $N$ states \textit{at the same time}. This corresponds to creating a superposition in the space register. Then, with an initialization procedure, we entangle each state of the space register to the corresponding state of the cell register, using CBE. The state of the lattice is as follows
\begin{equation} \label{eq:sublinear encoding}
    \ket{\Psi} = \frac{1}{\sqrt{N}} \sum_{x=0}^N\ket{x}\ket{\psi(x)} = \frac{1}{\sqrt{N}} \sum_{x=0}^N\ket{x}\ket{n_v(x) \dots n_0(x)}
\end{equation}
We can now carry out the collision on the cell register \textit{once}, and it will affect \textit{each cell}. This is called \textit{quantum parallelism} \cite{nielsen2010quantum}, and can lower the cost of the operations needed and of the computational resources dramatically. However, some problems arise. With the advantageous encoding in Eq.\ref{eq:sublinear encoding}, we cannot perform streaming and collision as we need. This is rooted in the following assumptions:
\begin{itemize}
    \item occupation states of the classical cell correspond to orthogonal states in the quantum register: this ensures the unitarity of the collision;
    \item the streaming operation is performed with a quantum walk procedure as in \cite{budinski2021quantum, budinski2021quantum2};
    \item  Different velocities must be represented in distinguishable states in the cell register, in order to apply a controlled shift for moving information in the correct direction.
\end{itemize}
Under these assumptions, coming from transposing the classical algorithm in quantum terms, it is not possible to perform unitary collision and streaming for multi-time step implementation. The full proof can be found in \ref{app2}, and expands the limitations already outlined in \cite{schalkers2024importance}, where an analogous proof was given for a specific encoding. We extend that proof to any possible encoding of $\ket{\psi(x)}$ in Eq.\ref{eq:sublinear encoding}, going beyond CBE. 

As discussed in the introduction, this is not proof that a quantum advantage cannot be obtained using CBE. It rather highlights some limitations for paving the way to new algorithms. The methods proposed in the following apply to any algorithm that includes the first assumption. The 2nd and 3rd assumptions can be adapted and changed for further research.

\subsection{\label{subsec:coll_circ} Collisional quantum circuits}
Considering CBE, we give here the general methods adopted for optimal decomposition of the collision operator into a series of quantum gates, being this a challenge for optimisation of QC algorithms.
The key idea is to use one or more ancillary qubits to be flipped if the cell is in a collisional configuration. Then, we apply a controlled collision with target the cell register, as shown in Fig. \ref{fig:gen_coll_circ}. 
\begin{figure}[ht]

\[
\begin{array}{c c}
\Qcircuit @C=1em @R=1em {
    \lstick{\ket{a}} & \ghost{V} & \ctrl{1} & \qw \\
    \lstick{\ket{\psi}} & \multigate{-1}{\hat{V}} & \gate{\hat{C}} & \qw
}
\end{array}
\]
\caption{$\ket{\psi}$ is the cell register, $\ket{a}$ is the ancillary register, $\hat{V}$ is the verification procedure, flipping the ancillary register only if there is a collisional state, $\hat{C}$ is the controlled collision applied to the cell register} \label{fig:gen_coll_circ}
\end{figure}

For D1Q3 we use two ancillary qubits for a QPE procedure with some specific operators, which identify a specific set of states. This is an example on how we can use QPE for the verification procedure. 

For FHP we leverage equivalence classes and logical exclusions, considerably lowering the decomposition cost of the collision operator. We implement 2-,3- and 4-body collisions through rotations of the cell, and the verification procedure is made with logical or equivalence class criteria. This gives the first proposed optimal overall circuit for FHP, allowing for a quantum simulation of FHP using CBE. The idea for carrying out non-deterministic collisions is to create a superposition of the two different outcomes and then make the system collapse using a measurement. This is reminiscent of a random extraction. With this method, we have different options, depending on the measurement we make. The first option is to create a superposition directly in the cell register. This means that if classically $s_0$ scatters into $s_1$ or $s_2$ with $50\%$ probability, the collision does the following
\begin{equation}
    \hat{C}\ket{s_i} = \ket{s_i'} = \frac{\ket{s_j}+\ket{s_k}}{\sqrt{2}}
\end{equation}
With $i,j,k \in \{0,1,2\}$ and different from each other. It is easy to prove that this operation is not unitary, since $\bra{s_i'}\ket{s_j'} \neq \bra{s_i}\ket{s_j} \forall i,j$. The alternative we are going to consider for calculating quantum invariants is to slightly change the classical collision, introducing the (low) probability of no collision. This means that the collision does the following
\begin{equation} \label{eq:unitary superpos}
    \hat{C} \ket{s_i} = \frac{-\ket{s_i} + 2 \ket{s_j} + 2 \ket{s_k}}{3}
\end{equation}
This operation is unitary. This fact will be used in Sec.\ref{subsec:fhp qinvariants}. If you use this unitary collision, you can measure one of the qubits and the state collapses into one of the possibilities, analogously to a random extraction. This carries out the desired collision and is used for studying the quantum invariants of the model.

The alternative we use in Sec.\ref{subsec:fhp coll circ}, more feasible as an algorithmic procedure, is to add an ancillary qubit $\ket{a}$ initialized to $\ket{0}$, thus performing the following operation
\begin{equation}
    \hat{C}\ket{s_i}\ket{a} = \frac{\ket{s_{i+1}}\ket{0}+\ket{s_{i+2}}\ket{1}}{\sqrt{2}}
\end{equation}
Where $i+1=\text{mod}_3(i+1)$ and $i+2=\text{mod}_3(i+2)$. This corresponds to the 2- and 4-body collisions of FHP. Using this method, we can measure the ancillary qubit obtaining the random extraction. This is the operation we are going to decompose optimally for the quantum collisional circuit of FHP. Measuring an ancilla does not solve the problem of reinitialization in case of sublinear encoding of the space, but offers a convenient decomposition of the unitary collision.

\subsection{\label{subsec:quantum_invariants} Quantum invariants}

We look at the conservation of local quantities as defined by Love~\cite{love2019quantum}, such as mass and momentum. Classically, for the discretization of the space, some nonphysical invariants arise~\cite{bernardin1992global}. The purpose of many studies has been to get rid of them for getting better results. This is necessary since any additional invariant can bring to a conservation equation that couples with mass and momentum conservation, vanishing the simulation purpose of the algorithm. In our quantum computing framework we can calculate the exact number of invariants, differently from the classical case.

As said in the introduction, the quantum counterpart of classical quantities, such as mass and momentum, are \textit{quantum observables}, which are Hermitian operators. An operator $\hat{O}$ is Hermitian if $\hat{O}^{\dagger}=\hat{O}$, where $\hat{O}^{\dagger}$ is the transpose conjugate of $\hat{O}$. Hermitian operators can be expressed as a linear combination of the Pauli operators, that span their orthonormal basis. For 1-qubit-observables these are $\mathcal{A}_1=\{I,X,Y,Z\}$ where
\begin{align}
    I=\begin{pmatrix}
        1 & 0 \\
        0 & 1
    \end{pmatrix} &&
    X=\begin{pmatrix}
        0 & 1 \\
        1 & 0
    \end{pmatrix} &&
    Y=\begin{pmatrix}
        0 & -i \\
        i & 0
    \end{pmatrix} &&
    Z=\begin{pmatrix}
        1 & 0 \\
        0 & -1
    \end{pmatrix}
\end{align}
If we consider 2-qubit-observables, their orthonormal basis is given by the cross product of the basis of the 1-qubit-observables $\mathcal{A}_2=\{II, IX, IY, IZ, $ $ XX, XY, XZ, YY, YZ, ZZ\}$ where the implicit operation is the cross product, i.e. $XX=X \otimes X$. This holds also for bigger dimensions, thus generally, a basis for $v$-qubit-observables $\mathcal{A}_v$ has $4^v$ elements, so that any quantum $v$-qubit-observable $\hat{O}$, i.e. any $2^v \cross 2^v$ Hermitian matrix, is
\begin{equation}
    \hat{O} = \sum_{i=0}^{4^v-1} \alpha_i \hat{O}_i
\end{equation}
where $\hat{O}_i \in \mathcal{A}_v$, and $\alpha_i$ are the corresponding coefficients.

How do observables, i.e. counterparts of classical quantities, evolve? If the state $\ket{\psi}$ undergoes an evolution $\hat{C}$, the evolved state is $\ket{\psi'}=\hat{C}\ket{\psi}$. Analogously, the evolved observables can be written as $\hat{O}' = \hat{C}^\dagger \hat{O} \hat{C}$.
Considering a cell with $v$ qubits, an observable $\hat{O}$ is a \textit{quantum invariant} if it commutes with the collision operator $\hat{C}$, so if it satisfies $\comm{\hat{C}}{\hat{O}}= \hat{C}\hat{O} - \hat{O} \hat{C} = 0$. We can write this property as follows, considering that for the unitarity of collision we have $\hat{C}\hat{C}^{\dagger}=\hat{C}^{\dagger}\hat{C}=I$, 

\begin{equation} \label{eq:commuting}
    \hat{C}^{\dagger} \hat{O} \hat{C} = \hat{O}    
\end{equation}

If we look at the lhs as the evolved operator, we can clearly see the \textit{invariant} property: the post-collision observable (lhs) is equal to the pre-collision observable (rhs). We can say that \textit{any} operator $\hat{O}$ respecting Eq.\ref{eq:commuting} is a quantum invariant. If we want to know the number of linearly independent quantum invariants  for a QLGCA model, we look at the evolution of each basis element $\hat{O}_i \in \mathcal{A}_v$. The number of quantum invariants corresponds to the number of linearly independent solutions of the linear system where each equation is the conservation equation Eq.\ref{eq:commuting} for $\hat{O}_i\in \mathcal{A}_v$. This translates into the following property

\begin{theorem}
Consider a collision of a quantum \dnqv model as a unitary operator $\hat{C}$. The number of linearly independent quantum invariants of a \dnqv model is $l=4^v-r$, where $r$ is the rank of the evolution matrix $M$, with elements
\begin{equation}
    M_{i,j}=\beta_{i,j}-\delta_{i,j}    
\end{equation}
where $\delta_{i,j}$ is the Kronecker delta and $\beta_{i,j}$ is the coefficient of the $\hat{O}_j$ operator for the Pauli decomposition of the evolved $\hat{O}_i$ operator, explicitly
$$ \hat{C}^\dagger \hat{O}_i \hat{C} = \sum_{j=0}^{4^v-1}\beta_{i,j} \hat{O}_j$$
for $\hat{O}_i\in \mathcal{A}_v$.
\end{theorem}
This methodology is applied to D1Q3 and FHP and confirms the expected conservation of mass and momentum operators, prooving the existence of several more and unexpected quantum invariants.

\subsection{\label{subsec: QPE} Quantum Phase Estimation}

The QPE algorithm is one of the standard subroutines in QC \cite{nielsen2010quantum}. Consider a unitary operator $\hat{U}_q$ with eigenvector $\ket{s}$ and corresponding eigenvalue $e^{i 2 \pi q(s)}$. We start with $n$ qubits in an additional register initialized to $\ket{0}$. These ancillae are controls of controlled-$\hat{U}_q$ operators, being $\ket{s}$ the target register, as shown in Fig.\ref{fig:qft_circ}.
\begin{figure}[ht]
\[
\begin{array}{c}

\Qcircuit @C=1.0em @R=0.2em @!R { \\
	 	\lstick{\ket{q_2}} & \qw & \gate{H} & \qw & \qw & \ctrl{3} & \multigate{2}{FT^\dagger} & \meter \\
	 	\lstick{\ket{q_1}} & \qw & \gate{H} & \qw & \ctrl{2} & \qw & \ghost{FT^\dagger} & \meter\\
	 	\lstick{\ket{q_0}} & \qw & \gate{H} & \ctrl{1} & \qw & \qw & \ghost{FT^\dagger} & \meter\\
	 	\lstick{\ket{s}} & \qw & \qw & \gate{U_q} & \gate{U_q^2} & \gate{U_q^4} & \qw & \qw\\
}
\end{array}
\]
\caption{Phase estimation algorithm for mass detection. $\ket{s}$ is the cell register, while $\ket{q_2q_1q_0}$ results in the approximated binary value of the quantity detected. $FT^\dagger$ corresponds to the inverse Fourier Transform} \label{fig:qft_circ}
\end{figure}
The state of the ancillary register acquires a relative phase according to $\hat{U}_q$ and $\ket{s}$. In the end, using the inverse quantum Fourier transform (IQFT), this relative phase translates into a binary approximation of the eigenvalue corresponding to $s$. Schematically, it provides the following operation

\begin{equation}
    \ket{0\dots 0}\ket{s} \longrightarrow \ket{\tilde{q}(s)} \ket{s}
\end{equation}
Where $\tilde{q}(s)$ is an approximation of $q(s)$. We can look at this procedure as a way of getting the information $q(s)$ about the state $\ket{s}$ depending on the operator $\hat{U}_q$. The idea underlying the method we propose is to define a diagonal operator $\hat{U}_q$ with elements $u_q(s,s') = \delta_{s,s'} e^{i q(s)}$ for $s,s'=0,\dots,2^v$ for a classical quantity $q$. An example of a mass operator used for QPE is given in Eq.\ref{eq:qpe matrix d1q3}. Being diagonal, the eigenstates of $\hat{U}_q$ correspond to the states of the cell adopting CBE. Thus, we can define the eigenvalues $e^{i q(s)}$ according to $s$, i.e. the state of the cell. In Eq.\ref{eq:qpe matrix d1q3} we see that states with the same mass have the same eigenvalue, so that when carrying out QPE we obtain the same result on the ancillary register. Another method for defining operators that are suitable for a QPE procedure involves the imaginary exponentiation of a quantum observable. If we consider an observable $\hat{Q}$, we can run the QPE using $\hat{U}_Q=e^{i\hat{Q}}$. Given the correspondence between classical quantities and quantum observables already outlined, this is always possible. For this paper, we preferred to use the first method as it gives more precise spectra for different values of the same quantity, as we see in the following sections.

Moving relevant physical information to an additional register allows us to measure the additional register instead of the cell register. For a linear encoding of the space, this avoids reinitialization in the case it is needed. For a sublinear encoding this QPE procedure does not solve the reinitialization obstacle and does not allow for a multi-time step implementation. However, our method allows for accessing information \textit{during the computation}: conditional operations for which a quantity is needed can be carried out, suggesting new optimizations. For QLBM a QPE procedure was proposed~\cite{kocherla2024fully}, but used in a different way. In this other work, since the information on the probability ditributions is stored in relative phases, QPE was conceived for direct information retrieving of the distributions of the cell.

In general, CBE allows us to retrieve quantities of interest with an additional register also with arithmetic operations, thus our method is not the only one for such a purpose. However, the method we propose can be applied to \textit{any} quantity, going beyond the ones accessible with arithmetic operations. This is of interest for considering a change of encoding or an investigation for non-classical quantities, that raise interest following the quantum invariants we outlined in the previous subsection. Thus, our method proposes new tools for future research, and is validated on D1Q3 and FHP.

\section{ \label{sec:d1q3} Application to the D1Q3 model}

\subsection{\label{subsec:d1q3 coll circuit}Collision circuit}

The collision exchanges a rest particle and 2 moving particles, doing $[010]\leftrightarrow[101]$. As we said, CBE consists of seeing the occupational states as a set of orthogonal quantum states. Thus, the collision can be represented as a unitary matrix
\begin{equation} \label{{eq:coll_matrix}}
    \hat{C} = 
    \begin{pmatrix}
        1 & 0 & 0 & 0 & 0 & 0 & 0 & 0 \\
        0 & 1 & 0 & 0 & 0 & 0 & 0 & 0 \\
        0 & 0 & 0 & 0 & 0 & 1 & 0 & 0 \\
        0 & 0 & 0 & 1 & 0 & 0 & 0 & 0 \\
        0 & 0 & 0 & 0 & 1 & 0 & 0 & 0 \\
        0 & 0 & 1 & 0 & 0 & 0 & 0 & 0 \\
        0 & 0 & 0 & 0 & 0 & 0 & 1 & 0 \\
        0 & 0 & 0 & 0 & 0 & 0 & 0 & 1
    \end{pmatrix}
\end{equation}
A gate-based implementation of this is given in~\cite{love2019quantum}. Even if Love's implementation is already optimal, we show that it is also possible to use the method in Sec.\ref{subsec:coll_circ} providing the verification procedure with QPE. We start from the operators $ZIZ$ and $IZZ$. If we look at their eigenvalues, practicing a QPE allows us to identify specific configurations. Dividing the space of possibilities again and again as shown in Fig. \ref{fig:tree_scheme_qpe_d1q3}, we arrive in the end to have a subset of the collisional state, $\{010,101\}$ in our case.
\begin{figure}[ht]
    \centering
    \begin{tikzpicture}
    \node at (0, 0) {$\{ 000,001,010,011,100,101,110,111 \}$};
    \node at (-1.2,-1) {$0$};
    \node at (1.2,-1) {$1$};
    \node at (-5,-1) {$ZIZ$};
    \node at (-5,-1.5) {eigenvalues};
    \draw[-stealth, black, thick] (0,-0.5) -- (2,-2);
    \draw[-stealth, black, thick] (0,-0.5) -- (-2,-2);
    \node at (2, -2.5) {$\{ 001,011,100,110 \}$};
    \node at (-2, -2.5) {$\{ 000,010,101,111 \}$};
    \node at (-5,-3.5) {$ZZI$};
    \node at (-5,-4) {eigenvalues};
    \draw[-stealth, black, thick] (2,-3) -- (3,-4.5);
    \draw[-stealth, black, thick] (2,-3) -- (1,-4.5);
    \node at (-3,-3.75) {$0$};
    \node at (-1,-3.75) {$1$};
    \draw[-stealth, black, thick] (-2,-3) -- (-3,-4.5);
    \draw[-stealth, black, thick] (-2,-3) -- (-1,-4.5);
    \node at (+3,-3.75) {$1$};
    \node at (+1,-3.75) {$0$};
    \node at (-3, -5) {$\{ 000, 111\}$};
    \node at (-1, -5) {$\{ 010,101 \}$};
    \node at (1, -5) {$\{ 001,110 \}$};
    \node at (3, -5) {$\{ 011,100 \}$};
    \end{tikzpicture}
    \caption{Tree scheme to see the partition of states depending on the detected eigenvalue of associated operator for QPE.}
    \label{fig:tree_scheme_qpe_d1q3}
\end{figure}
We can call $ZIZ$ and $ZZI$ the \textit{discrimination operators}, and apply a QPE scheme as shown in Fig.\ref{fig:pe_coll}. The operators were chosen to look at the different possible eigenvalues of operators on 3 qubits involving $Z$-gates. Operators involving $X$-gates and $Y$-gates were excluded from the CBE adopted. The collision circuit for D1Q3 in Fig.\ref{fig:pe_coll} represents a novel method for executing collisions.

\begin{figure}[ht]
\[
\begin{array}{c}

\Qcircuit @C=1.0em @R=0.2em @!R { \\ & & &                        \mbox{QPE Verification} & & & & & & \mbox{Collision} & & & & &\\ \\
	 	\lstick{\ket{z_1}} & \gate{H} & \qw & \qw & \ctrl{3} & \ctrl{2} & \gate{H} & \qw & \ctrl{1} & \ctrl{1} & \ctrl{1} & \qw\\ 
	 	\lstick{\ket{z_0}} & \gate{H} & \ctrl{1} & \ctrl{3} & \qw & \qw & \gate{H} & \qw & \ctrlo{1} & \ctrlo{2} & \ctrlo{3} & \qw\\
	 	\lstick{\ket{n_2}} & \qw & \gate{Z} & \qw & \qw & \gate{Z} & \qw & \qw & \targ & \qw & \qw & \qw\\
	 	\lstick{\ket{n_1}} & \qw & \qw & \qw & \gate{Z} & \qw & \qw & \qw & \qw & \targ & \qw & \qw\\
	 	\lstick{\ket{n_0}} & \qw & \qw & \gate{Z} & \qw & \qw & \qw & \qw & \qw & \qw & \targ & \qw \gategroup{4}{2}{8}{7}{1.5em}{--} \gategroup{4}{9}{8}{11}{1.5em}{--}\\
}
\end{array}
\]
\caption{ Collision circuit for D1Q3. The first two qubits are conditional: the practicing of the collision, which is a series of Toffoli, depends on their state. This is a phase estimation algorithm with the first two as additional registers. The result, for the simplicity of the model, is deterministic.} \label{fig:pe_coll}
\end{figure}

\subsection{\label{subsec:d1q3 qinvariants}Quantum invariants}

The first thing to do for calculating the quantum invariants is writing the collision operator. The collision operator of D1Q3 presented in~\cite{love2019quantum} can be written as

\begin{equation} \label{eq:coll d1q3} \begin{split}
    \hat{C}_{tot}=\frac{1}{4}(& 3III+IZZ+XXX+XYY+ \\ & -YXY+YYX-ZIZ+ZZI)
    \end{split}
\end{equation}

The second step is to calculate the evolution of each Pauli operator on 3 qubit space. The results are in Table \ref{tab:d1q3_evolution}. From this it is possible to verify the conservation of mass $m$ and momentum $p$ as defined by Love :

\begin{equation}
    m = IIZ + 2IZI + ZII 
\end{equation}
\begin{equation}
    p = IIZ - ZII    
\end{equation}

Looking at all the evolved operators we see that there are more conserved quantities than expected. We can compute their number by calculating the evolution matrix and its rank, as explained in the Sec.\ref{subsec:quantum_invariants}. For D1Q3 the rank is equal to 14, meaning that there are 50 linearly independent conserved quantities.

\begin{table}[ht]
    \centering
    \begin{tabular}{c|c}
        Classical counterpart & Symmetric counterpart \\ \hline
        $III$ & $XXX$\\
        $IZZ,ZIZ,ZZI$ & $XYY,YXY,YYX$\\
        $m = IIZ + 2IZI + ZII$ & $XXY + 2YXX + YXX$\\
        $p = IIZ - ZII$ & $XXY - YXX$\\
        & $IXI+XIX$
    \end{tabular}
    \caption{We can see that the quantum formulation introduces a symmetry on conserved quantities. This happens also in FHP}
    \label{tab:my_label}
\end{table}

The conservation of all these quantities is unexpected. The quantum invariants $IZZ,ZIZ,ZZI$ have explainable classical counterparts as $n_0 \oplus n_1$, $n_0 \oplus n_2$, $n_1 \oplus n_2$. This has a clear explanation in the CBE adopted. The states $\ket{0}$ and $\ket{1}$ are the eigenstates of $Z$ operator. Thus, the $Z$ operator is reminiscent of the presence of a particle. Others are due to a symmetry of the collision operator with respect to a change of basis, as suggested in Table\ref{tab:my_label}. However, the remaining ones are unexpected, and a clear and unique correspondence between quantum and classical quantities is left as a future perspective. Our finding is intended to highlight this feature of QC algorithm for LGA that has never been considered in such detail before. 

\subsection{\label{subsec:d1q3 qpe}Quantum phase estimation for quantities}

The methodology consists of defining the quantum operator to have the desired eigenvalues for a phase estimation procedure, as explained in the methods section. For example, the matrix representation of the mass operator is the following

\begin{equation} \label{eq:qpe matrix d1q3}
    \hat{U}_m = \begin{pmatrix}
        1 & 0 & 0 & 0 & 0 & 0 & 0 & 0 \\
        0 & e^{-i} & 0 & 0 & 0 & 0 & 0 & 0 \\
        0 & 0 & e^{-2i} & 0 & 0 & 0 & 0 & 0 \\
        0 & 0 & 0 & e^{-3i} & 0 & 0 & 0 & 0 \\
        0 & 0 & 0 & 0 & e^{-i} & 0 & 0 & 0 \\
        0 & 0 & 0 & 0 & 0 & e^{-2i} & 0 & 0 \\
        0 & 0 & 0 & 0 & 0 & 0 & e^{-3i} & 0 \\
        0 & 0 & 0 & 0 & 0 & 0 & 0 & e^{-4i} \\
    \end{pmatrix}
\end{equation}

\begin{figure}[ht]
    \centering
    \includegraphics[scale=0.23]{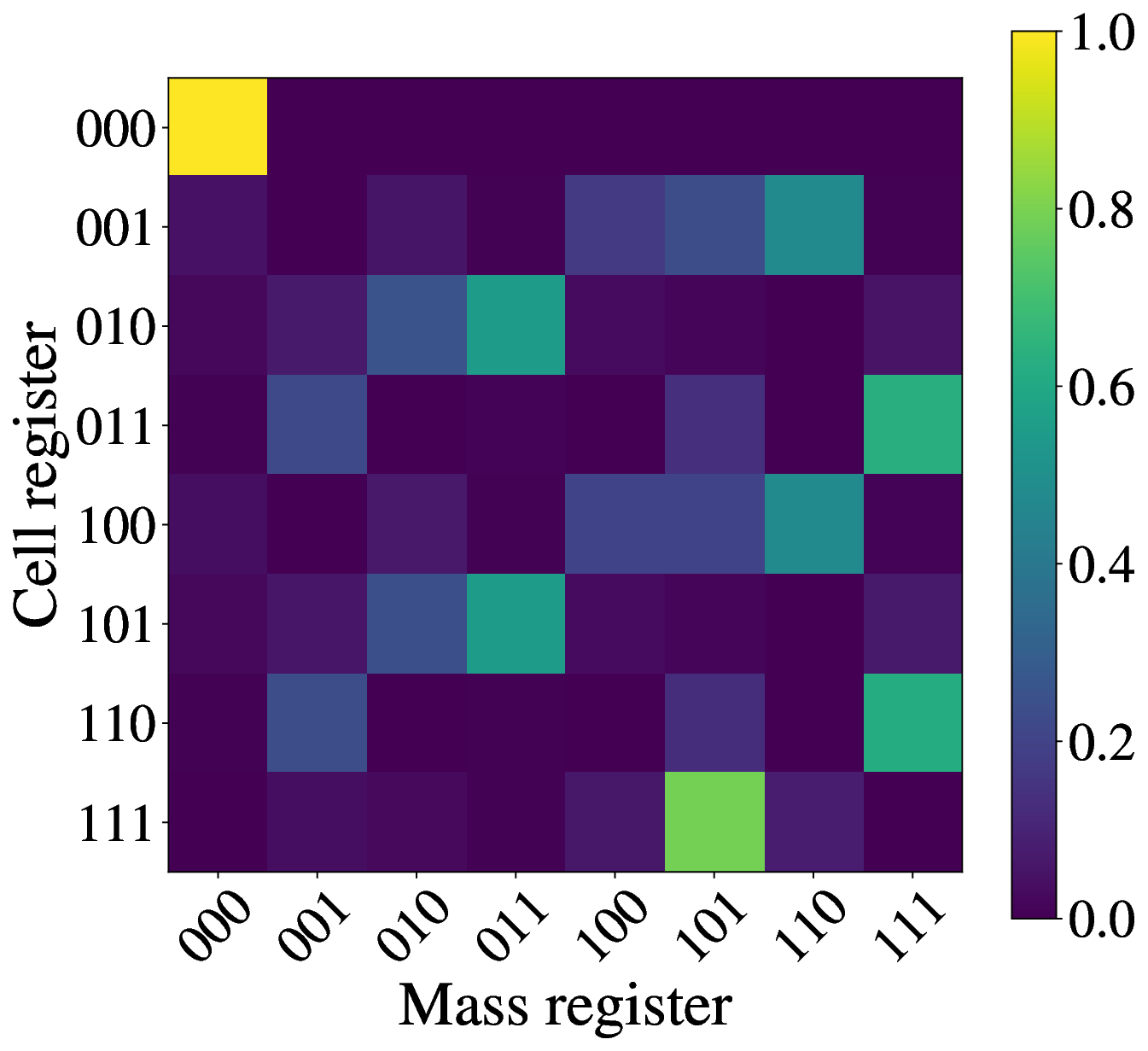}
    \includegraphics[scale=0.23]{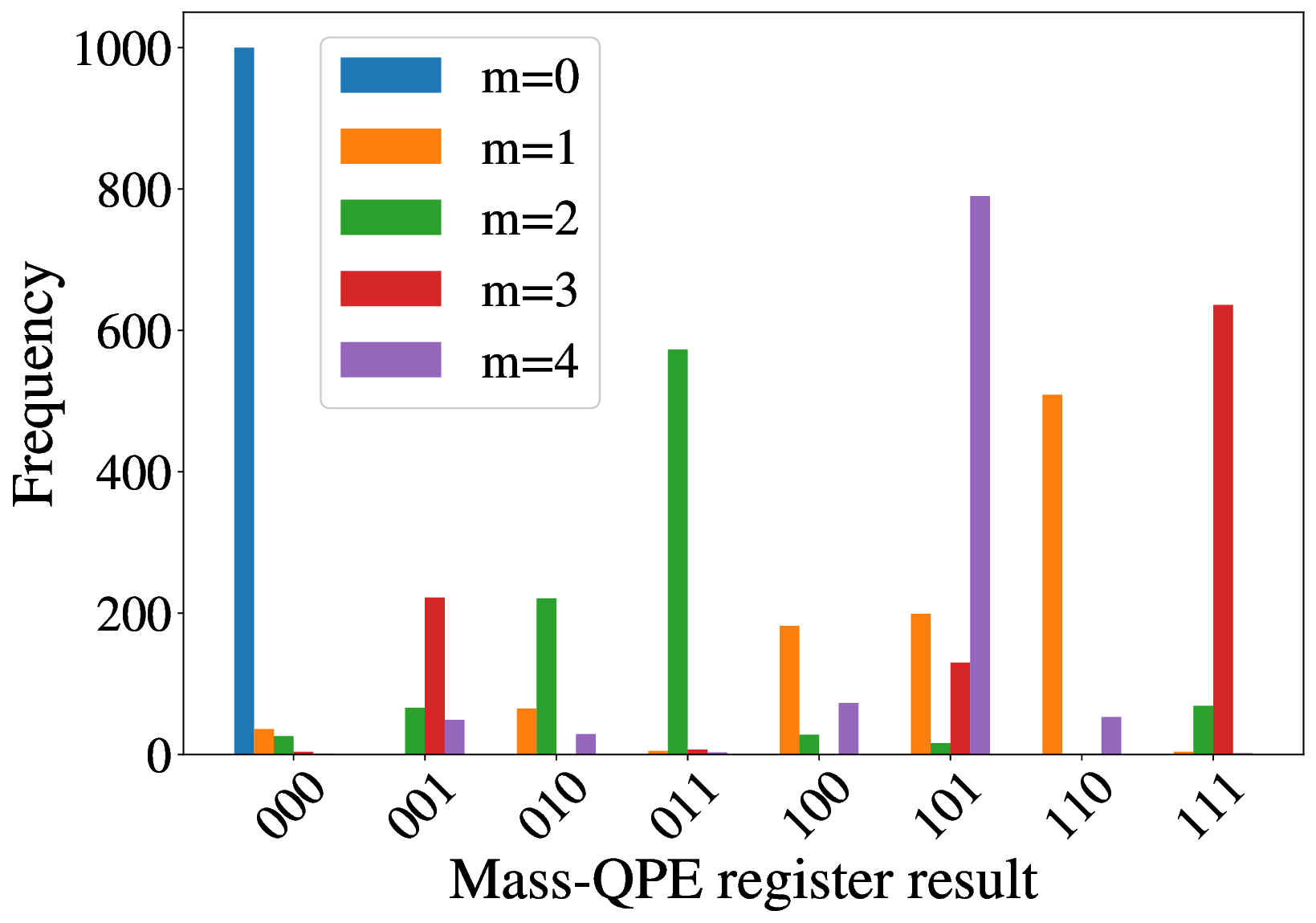}
    \caption{On the left: for each input state (y-axis) we plot the measurement outcome probability on the mass register (x-axis) carrying out the phase estimation algorithm in Fig. \ref{fig:qft_circ}. We can see the success of the QPE algorithm in identifying the same rows for states with the same quantities. On the right: measurement outcome frequencies depending explicitly on the mass. We can distinguish different peaks that identify different masses with different outputs in the QPE register. The probabilities reported come from the cell states  $\{000,001,101,011,111\}$}
    \label{fig:qpe_prob}
\end{figure}

We clearly see that the eigenvalues of this operator correspond to $e^{-im(s)}$ where $m(s)$ is the mass of the state $s$ (e.g. $m(\qty{001})=1$). This turns the problem of measuring the mass of the cell into a phase estimation problem. Using $\hat{U}_m$ as an operator for QPE the circuit in Fig.\ref{fig:qft_circ}, we get the results in Fig. \ref{fig:qpe_prob}.
We see that the rows of states $\ket{001},\ket{100}$ and $\ket{101},\ket{010}$, and $\ket{110},\ket{011}$ are the same. This means that the algorithm can detect the same eigenvalues, i.e. the same mass. This procedure can be seen as a possible quantum implementation of \texttt{get\_mass()} and \texttt{get\_momentum()} functions in Algorithm\ref{alg:cap}, that is probabilistic but manages to avoid direct measurement of the cell, and can be applied to any quantity of interest. The best result that can be obtained is having one peak for each eigenvalue, i.e. for each classical value of the corresponding quantity. This optimization, left as a future perspective, can allow to calculate directly quantities that can be used for other quantum subroutines.

\section{\label{sec:fhp} Application to the FHP LGCA  model}

\subsection{\label{subsec:fhp coll circ}Collision circuit}

We have $v$ qubits representing the cell, these compose the \textit{cell register}. 
Each collision is a rotation in the cell, as represented in Table \ref{tab:collisions2}. It is possible to define these rotations in terms of quantum operations, as a series of swaps gates in Figure\ref{fig:rotations}. 

\begin{figure}[ht]

\[
\begin{array}{c c}

\Qcircuit @C=1em @R=1em {
& & \mbox{180°} & & & & & &
& & \mbox{120°} & & & & &
& & & \mbox{60°} \\
\lstick{\ket{n_5}} 
& \qw & \qw & \qswap & \qw & & &
& \qswap & \qswap & \qw & \qw & \qw & & &
& \qswap & \qswap & \qswap & \qswap & \qswap & \qw\\
\lstick{\ket{n_4}} 
& \qw & \qswap & \qw \qwx & \qw & & &
& \qw \qwx & \qw \qwx & \qswap & \qswap & \qw & & &
& \qswap \qwx & \qw \qwx & \qw \qwx & \qw \qwx & \qw \qwx & \qw\\
\lstick{\ket{n_3}} 
& \qswap & \qw \qwx & \qw \qwx & \qw & & &
& \qswap \qwx & \qw \qwx & \qw \qwx & \qw \qwx & \qw & & & 
& \qw & \qswap \qwx & \qw \qwx & \qw \qwx & \qw \qwx & \qw\\
\lstick{\ket{n_2}} 
& \qw \qwx & \qw \qwx & \qswap \qwx & \qw & & &
& \qw & \qw \qwx & \qswap \qwx & \qw \qwx & \qw & & &
& \qw & \qw & \qswap \qwx & \qw \qwx & \qw \qwx & \qw\\
\lstick{\ket{n_1}} 
& \qw \qwx & \qswap \qwx & \qw & \qw & & &
& \qw & \qswap \qwx & \qw & \qw \qwx & \qw & & &
& \qw & \qw & \qw & \qswap \qwx & \qw \qwx & \qw\\
\lstick{\ket{n_0}} 
& \qswap \qwx & \qw & \qw & \qw & & &
& \qw & \qw & \qw & \qswap \qwx & \qw & & &
& \qw & \qw & \qw & \qw & \qswap \qwx & \qw
}

\end{array}
\]
\caption{Rotations with quantum circuits} \label{fig:rotations}
\end{figure}

Applying the method explained in Sec.\ref{subsec:coll_circ}, we develop an optimal overall circuit for the implementation of 0-momentum collisions in Table\ref{tab:collisions2}, as shown in Figure \ref{fig:b234}. Each collision is implemented in two parts. We have first the verification of the collisional states that uses one conditional qubit $\ket{b}$.The second part is the collision, which is carried out as a controlled operation on the conditional qubit using an ancilla $\ket{a}$ for simulating random outcomes of $B_2$ and $B_4$. 

\begin{figure*}[ht]

\[
\begin{array}{c}
\scalebox{0.5}{
\Qcircuit @C=1.0em @R=0.2em @!R { \\
        \\ & & & & & & \mbox{\Large$B_3$-verification} & & & & & & & & \mbox{\Large $B_3$} & & & & & & \mbox{\Large $B_{2,4}$-verification} & & & & & & & & & & \mbox{\Large $B_{2,4}$}  \\ \\
	 	\lstick{\ket{n_5}} & \targ & \qw & \qw & \qw & \qw & \ctrl{1} & \qw & \qw & \qw & \qw & \targ & \qw & \qw & \qw & \qswap \qwx[3] & \qw & \qw & \qw & \targ & \ctrlo{1} & \targ & \qw & \qw & \qw & \qswap & \qswap & \qw & \qw & \qw & \qswap & \qswap & \qw & \qw & \qw & \qw  \\
	 	\lstick{\ket{n_4}} & \ctrl{-1} & \targ & \qw & \qw & \qw & \ctrl{1} & \qw & \qw & \qw & \targ & \ctrl{-1} & \qw & \qw & \qswap \qwx[3] & \qw & \qw & \qw & \targ & \qw & \ctrlo{1} & \qw & \targ & \qw & \qw & \qw & \qw & \qswap & \qswap & \qw & \qw & \qw & \qswap & \qswap & \qw & \qw\\
	 	\lstick{\ket{n_3}} & \qw & \ctrl{-1} & \targ & \qw & \qw & \ctrl{1} & \qw & \qw & \targ & \ctrl{-1} & \qw & \qw & \qswap \qwx[3] & \qw & \qw & \qw & \targ & \qw & \qw & \ctrlo{4} & \qw & \qw & \targ & \qw & \qswap \qwx[-2] & \qw & \qw & \qw & \qw & \qswap \qwx[-2] & \qw & \qw & \qw & \qw & \qw\\
	 	\lstick{\ket{n_2}} & \qw & \qw & \ctrl{-1} & \targ & \qw & \ctrl{1} & \qw & \targ & \ctrl{-1} & \qw & \qw & \qw & \qw & \qw & \qswap & \qw & \qw & \qw & \ctrl{-3} & \qw & \ctrl{-3} & \qw & \qw & \qw & \qw & \qw & \qswap \qwx[-2] & \qw & \qw & \qw & \qw & \qswap \qwx[-2] & \qw & \qw & \qw\\
	 	\lstick{\ket{n_1}} & \qw & \qw & \qw & \ctrl{-1} & \targ & \ctrl{2} & \targ & \ctrl{-1} & \qw & \qw & \qw & \qw & \qw & \qswap & \qw & \qw & \qw & \ctrl{-3} & \qw & \qw & \qw & \ctrl{-3} & \qw & \qw & \qw & \qswap \qwx[-4] & \qw & \qw & \qw & \qw & \qswap \qwx[-4] & \qw & \qw & \qw & \qw\\
	 	\lstick{\ket{n_0}} & \qw & \qw & \qw & \qw & \ctrl{-1} & \qw & \ctrl{-1} & \qw & \qw & \qw & \qw & \qw & \qswap & \qw & \qw & \qw & \ctrl{-3} &  \qw & \qw & \qw & \qw & \qw & \ctrl{-3} & \qw & \qw & \qw & \qw & \qswap \qwx[-4] & \qw & \qw & \qw & \qw & \qswap \qwx[-4] & \qw & \qw\\
	 	\lstick{\ket{b}} & \qw & \qw & \qw & \qw & \qw & \targ & \qw & \qw & \qw & \qw & \qw & \qw & \ctrl{-1} & \ctrl{-2} & \ctrl{-3} & \qw & \qw & \qw & \qw & \targ & \qw & \qw & \qw & \qw & \ctrl{-4} & \ctrl{-2} & \ctrl{-3} & \ctrl{-1} & \ctrl{1} & \qw & \qw & \qw & \qw & \qw & \qw \\
        \lstick{\ket{a}} & \qw & \qw & \qw & \qw & \qw & \qw & \qw & \qw & \qw & \qw & \qw & \qw & \qw & \qw & \qw & \qw & \qw & \qw & \qw & \qw & \qw & \qw & \qw & \qw & \qw & \qw & \qw & \qw & \gate{H} & \ctrl{-5} & \ctrl{-3} & \ctrl{-4} & \ctrl{-2} & \meter & \qw  \gategroup{5}{2}{11}{12}{1.5em}{--} \gategroup{5}{14}{11}{16}{1.5em}{--} \gategroup{5}{18}{11}{24}{1.5em}{--} \gategroup{5}{26}{12}{35}{1.5em}{--}\\ 
\\          }}
\end{array}
\]
\caption{Collisional circuit for 0-momentum collisions of FHP. These are performed depending on a conditional qubit $\ket{b}$ that gets flipped if the input state is collisional, and an additional qubit $\ket{a}$ that introduces the non-deterministic character of 2- and 4-body collisions and is measured at each time-step. This circuit uses the invariance of collisional states for the merging of all the collisions in one circuit.} \label{fig:b234}
\end{figure*}

To implement the verification of collisional states of $B_3$ we consider a logic methodology, starting from the expression that can be found in~\cite{wolf2004lattice}, adapted to our convention.

\begin{equation}
    b = (n_0 \wedge n_1) \& (n_1 \wedge n_2) \& (n_2 \wedge n_3) \& (n_3 \wedge n_4) \& (n_4 \wedge n_5)    
\end{equation}

Where $\wedge$ is a XOR operation and $\&$ is an AND operation. We can turn this logic expression in a quantum circuit interpreting $\wedge$ as CNOT and $\&$ as a generalized Toffoli gate with the cell's qubits as controls and $\ket{b}$ as a target. Then we restore the original cell and we apply the C-Swap gates for a rotation of 180° with $\ket{b}$ as a control qubit. 

To implement the verification of collisional states of $B_2$ and $B_4$, we apply a reasoning based on equivalence classes. We define an \textit{asymmetric opposite pair} as a pair of opposite bit-velocities that differ from each other (e.g. $\ket{100000}$ has $1$ opposite pairs for the asymmetry between $\ket{n_5}$ and $\ket{n_2}$, $\ket{110000}$ has $2$ asymmetric opposite pairs for $\ket{n_5}$,$\ket{n_4}$ and $\ket{n_2}$,$\ket{n_1}$, $\ket{100100}$ has $0$ asymmetric opposite pairs). We consider the equivalence class of states with $0$ asymmetric opposite pairs represented in Table\ref{tab:0asymmetric}. 

\begin{table}[ht]
\centering
\begin{tabular}{| c | c c c |}
    \hline \hline
    Invariant states & \scalebox{0.60}{\begin{tikzpicture}
    \draw[-stealth, black, thick] (0.13,0) -- (1,0);
    \draw[-stealth, black, thick] (0.065,0.11258) -- (0.5,0.866);
    \draw[-stealth, black, thick] (-0.065,0.11258) -- (-0.5,0.866);
    \draw[-stealth, black, thick] (-0.13,0) -- (-1,0);
    \draw[-stealth, black, thick] (-0.065,-0.11258) -- (-0.5,-0.866);
    \draw[-stealth, black, thick] (0.065,-0.11258) -- (0.5,-0.866);
    \draw[black] (0,0) circle (4pt);
    \end{tikzpicture}} & \scalebox{0.60}{\begin{tikzpicture}
    \draw[-stealth, white, thick] (0.13,0) -- (1,0);
    \draw[-stealth, white, thick] (0.065,0.11258) -- (0.5,0.866);
    \draw[-stealth, white, thick] (-0.065,0.11258) -- (-0.5,0.866);
    \draw[-stealth, white, thick] (-0.13,0) -- (-1,0);
    \draw[-stealth, white, thick] (-0.065,-0.11258) -- (-0.5,-0.866);
    \draw[-stealth, white, thick] (0.065,-0.11258) -- (0.5,-0.866);
    \draw[black] (0,0) circle (4pt);
\end{tikzpicture}} &  \\
    $B_2$ collisional states & \scalebox{0.60}{\begin{tikzpicture}
    \draw[-stealth, black, thick] (0.13,0) -- (1,0);
    \draw[-stealth, white, thick] (0.065,0.11258) -- (0.5,0.866);
    \draw[-stealth, white, thick] (-0.065,0.11258) -- (-0.5,0.866);
    \draw[-stealth, black, thick] (-0.13,0) -- (-1,0);
    \draw[-stealth, white, thick] (-0.065,-0.11258) -- (-0.5,-0.866);
    \draw[-stealth, white, thick] (0.065,-0.11258) -- (0.5,-0.866);
    \draw[black] (0,0) circle (4pt);
\end{tikzpicture}}  & \scalebox{0.60}{\begin{tikzpicture}
    \draw[-stealth, white, thick] (0.13,0) -- (1,0);
    \draw[-stealth, black, thick] (0.065,0.11258) -- (0.5,0.866);
    \draw[-stealth, white, thick] (-0.065,0.11258) -- (-0.5,0.866);
    \draw[-stealth, white, thick] (-0.13,0) -- (-1,0);
    \draw[-stealth, black, thick] (-0.065,-0.11258) -- (-0.5,-0.866);
    \draw[-stealth, white, thick] (0.065,-0.11258) -- (0.5,-0.866);
    \draw[black] (0,0) circle (4pt);
\end{tikzpicture}} & \scalebox{0.60}{\begin{tikzpicture}
    \draw[-stealth, white, thick] (0.13,0) -- (1,0);
    \draw[-stealth, white, thick] (0.065,0.11258) -- (0.5,0.866);
    \draw[-stealth, black, thick] (-0.065,0.11258) -- (-0.5,0.866);
    \draw[-stealth, white, thick] (-0.13,0) -- (-1,0);
    \draw[-stealth, white, thick] (-0.065,-0.11258) -- (-0.5,-0.866);
    \draw[-stealth, black, thick] (0.065,-0.11258) -- (0.5,-0.866);
    \draw[black] (0,0) circle (4pt);
\end{tikzpicture}} \\
    $B_4$ collisional states & \scalebox{0.60}{\begin{tikzpicture}
    \draw[-stealth, black, thick] (0.13,0) -- (1,0);
    \draw[-stealth, black, thick] (0.065,0.11258) -- (0.5,0.866);
    \draw[-stealth, white, thick] (-0.065,0.11258) -- (-0.5,0.866);
    \draw[-stealth, black, thick] (-0.13,0) -- (-1,0);
    \draw[-stealth, black, thick] (-0.065,-0.11258) -- (-0.5,-0.866);
    \draw[-stealth, white, thick] (0.065,-0.11258) -- (0.5,-0.866);
    \draw[black] (0,0) circle (4pt);
\end{tikzpicture}} & \scalebox{0.60}{\begin{tikzpicture}
    \draw[-stealth, black, thick] (0.13,0) -- (1,0);
    \draw[-stealth, white, thick] (0.065,0.11258) -- (0.5,0.866);
    \draw[-stealth, black, thick] (-0.065,0.11258) -- (-0.5,0.866);
    \draw[-stealth, black, thick] (-0.13,0) -- (-1,0);
    \draw[-stealth, white, thick] (-0.065,-0.11258) -- (-0.5,-0.866);
    \draw[-stealth, black, thick] (0.065,-0.11258) -- (0.5,-0.866);
    \draw[black] (0,0) circle (4pt);
\end{tikzpicture}} & \scalebox{0.60}{\begin{tikzpicture}
    \draw[-stealth, white, thick] (0.13,0) -- (1,0);
    \draw[-stealth, black, thick] (0.065,0.11258) -- (0.5,0.866);
    \draw[-stealth, black, thick] (-0.065,0.11258) -- (-0.5,0.866);
    \draw[-stealth, white, thick] (-0.13,0) -- (-1,0);
    \draw[-stealth, black, thick] (-0.065,-0.11258) -- (-0.5,-0.866);
    \draw[-stealth, black, thick] (0.065,-0.11258) -- (0.5,-0.866);
    \draw[black] (0,0) circle (4pt);
\end{tikzpicture}} \\
    \hline \hline
\end{tabular}
\caption{\label{tab:0asymmetric}%
States with 0 asymmetric pairs. These states are invariant under a rotation of 180°, which corresponds to a $B_3$ collision.
}
\end{table}

\begin{figure}[ht]
    \centering
    \includegraphics[scale=0.5]{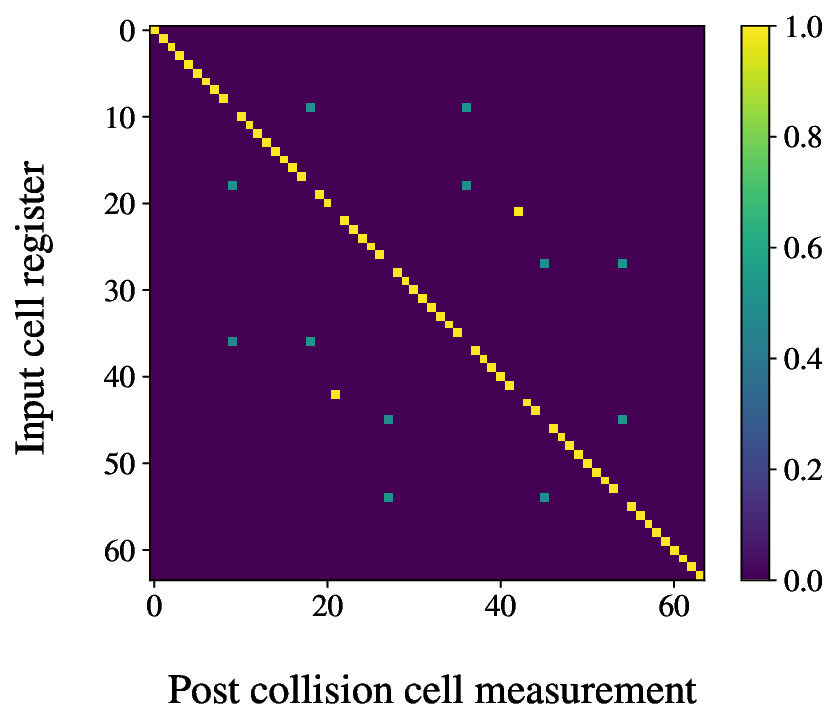}
    \caption{Probabilities of outcome from a measurement of the cell register at the end of the collisional circuit in Fig.\ref{fig:b234}. We can see that the collision operator applied is a diagonal operator except for $B_3$ collisional states (21,42), $B_2$ collisional states (9,18,36), $B_4$ collisional states (27,45,54). This gaurantees the correctness of the decomposition.}
    \label{fig:res_verification}
\end{figure}

The collisional states of $B_2$ and $B_4$ belong to this class, and the other states are invariant under these collisions and under $B_3$. Thus, we can target $\ket{b}$ to be $\ket{1}$ if there are no asymmetric opposite pairs. Then, with a series of C-Swaps with $\ket{b}$ as a control, we apply the first 120° rotation. We apply a controlled-H gate with $\ket{b}$ as control and $\ket{a}$, initialized to $\ket{0}$, as target. Then we apply the same controlled rotation of 120° with $\ket{a}$ as a control. In this way, the state $\ket{0}$ of $\ket{a}$ is entangled to a rotation of 120°, while the state $\ket{1}$ is entangled to a rotation of 240°, but only if $\ket{b}=\ket{1}$. Measuring $\ket{a}$ will cause a collapse of the cell state into one of the two, resembling a random extraction. The two collision can be made on the same circuit and with the same ancilla $\ket{b}$ for the invariances introduced in Sec.\ref{sec:classical models}. The circuit in Figure \ref{fig:b234} was verified in Qiskit applying a measurement on the cell register. The results are reported in Fig.\ref{fig:res_verification}.

The progress we report stands in the computational cost of this quantum procedure. Each algorithm must be decomposed in a set of universal gates to be executed on a quantum computer. The simulation tool we used, Qiskit\cite{javadi2024quantum}, provides a decomposition method for this purpose. The algorithm we developed was decomposed by Qiskit in 291 gates. If we try to decompose directly the collision operator that should give the same result as in Fig.\ref{fig:res_verification}, we get an order of $10^4$ gates. Our method shows an optimization over the deafult decomposition of Qiskit of $\approx 100$ times. We also propose to use for the first time the features of the classical configurations to find optimal quantum circuits, that can be further applied to other collisions with the use of more ancillary $\ket{b}$ qubits.

\subsection{\label{subsec:fhp qinvariants}Quantum invariants}

We apply the methodology of the evolution matrix $M$ for finding the number of quantum invariants. For FHP we need first to introduce random collisions. For calculating the number of quantum invariants, we need to write the unitary operator $\hat{C}$. We cannot take it directly from Figure \ref{fig:b234} because it involves ancillary qubits and applies a measurement, which makes the operator non-unitary. However, we can consider the analogous unitary operation that, instead of relying on an ancilla, creates a superposition of desired states.
This is the collision stated in Sec.\ref{subsec:coll_circ}, specifically in Eq.\ref{eq:unitary superpos}. This collision is not \textit{precisely} the one of FHP. However, it conserves mass and momentum, and it can still be used for seeing spurious quantum invariants. A method for calculating quantum invariants of non-unitary operator is left as a future perspective.

As we can see from Table \ref{tab:_res_M} we verified that the number of quantum invariants becomes smaller if we increase the number of collisions. This was a result expected classically.
\begin{table}[ht]
\centering
\begin{tabular}{l | c | c }
  Collisions & Rank of M & Quantum invariants \\ \hline
  $B_3$ & 126 & 3970 \\
  $B_{2,4}$ & 488 & 3608 \\
  $B_{2,3,4}$ & 590 & 3506 \\
\end{tabular}
\caption{Rank of evolution matrix corresponding to different collisions introduced in the unitary. The number of quantum invariants corresponds to $4^6-r$ where $r$ is the rank of the evolution matrix}\label{tab:_res_M}
\end{table}
Moreover, we verified the conservation of mass and momentum as the linear combination of Pauli operators defined by Love~\cite{love2019quantum}

\begin{eqnarray}
    M = Z_0 + Z_1 + Z_2 + Z_3 + Z_4 + Z_5 \\
    P_x = Z_0 - Z_3 + \frac{1}{2} (Z_1 + Z_5 - Z_2 - Z_4) \\
    P_y = \frac{\sqrt{3}}{2} ( Z_1 + Z_2 - Z_4 - Z_5) 
\end{eqnarray}

Where $Z_i$ is the Z operator on the i-th qubit (e.g. $Z_2 = IIIZII$,  $Z_{0,3}=IIZIIZ$). Beyond these operators, analogously to D1Q3, other operators with a classical counterpart are conserved, considering $B_{2,3,4}$, as $I^{\otimes 6}$, $Z_{0,3}$,$Z_{1,4}$,$Z_{0,1,3,4}$ etc., as well as the symmetric respect to the $X^{\otimes 6}$ multiplication. All these quantities are unexpected, but we should consider that in the end we are emulating exactly the classical system with its conservation laws. Thus, we believe that the measurement process makes these additional invariants negligible, and the resulting algorithm can be used for FHP simulation on a quantum computer. A more detailed analysis of the quantum invariants is out of the scope of this paper.

\begin{figure}[ht]
    \centering
    \includegraphics[scale=0.40]{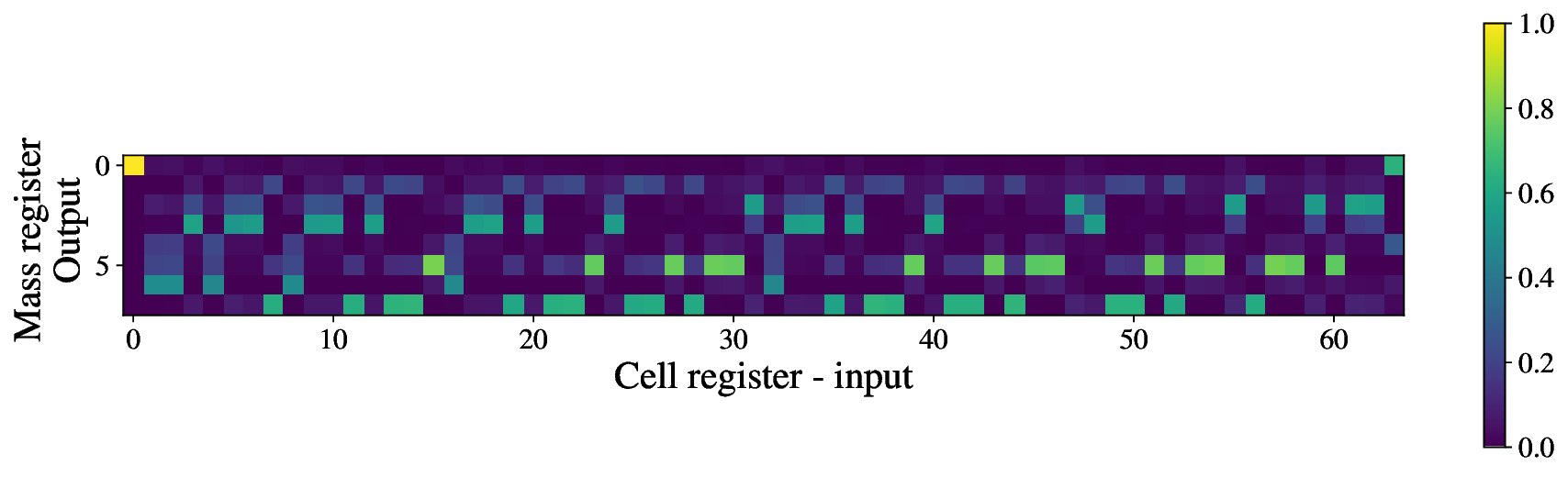}
    \includegraphics[scale=0.35]{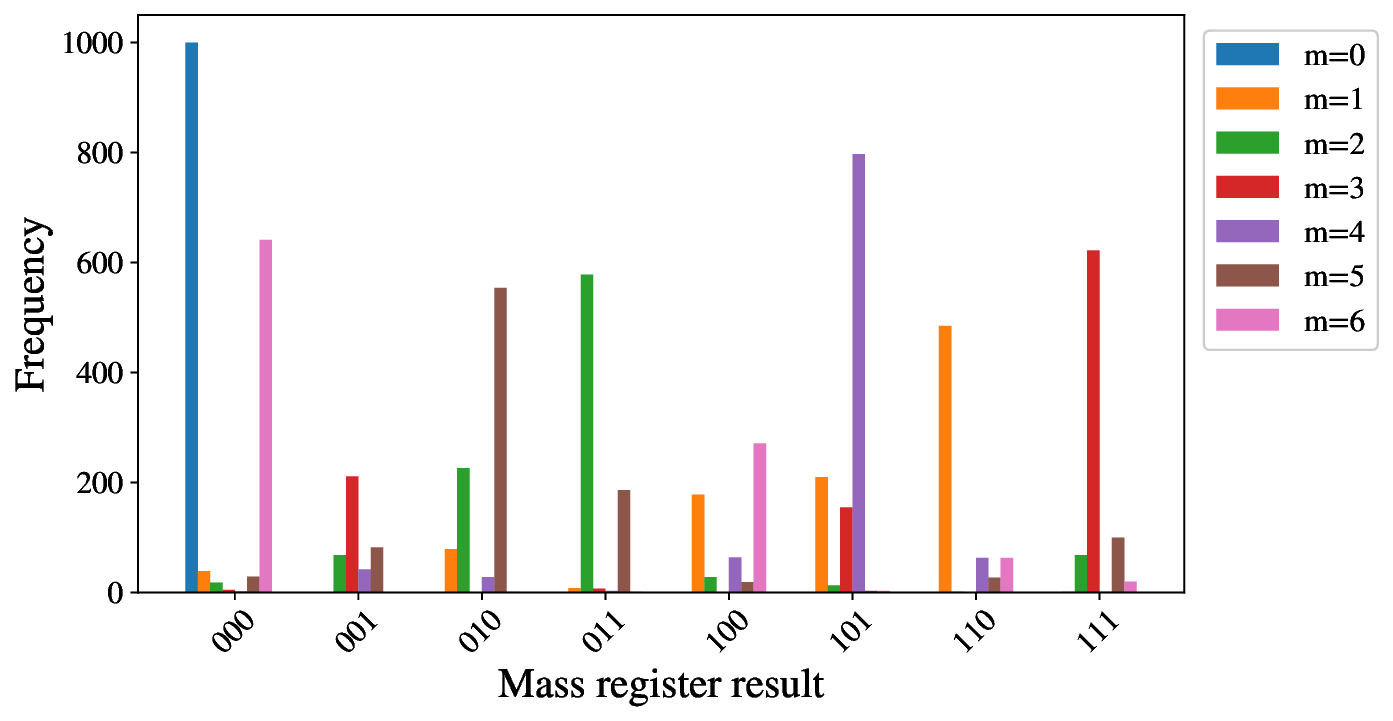}
    \caption{Above: for each input state (x-axis) we plot the measurement outcome probability on the mass register (y-axis) carrying out the phase estimation algorithm in Fig. \ref{fig:qft_circ}. We can see the success of the QPE algorithm in identifying the same columns for states with the same quantities. Below: measurement outcome frequencies depending explicitly on the mass. We can distinguish different peaks that identify different masses with different outputs in the QPE register. The probabilities reported come from the cell states  $\{2^i\}$ for $i\in \{0,1,2,3,4,5 \}$}
    \label{fig:qpe_fhp}
\end{figure}

\subsection{\label{subsec:fhp qpe}Quantum phase estimation for quantities}

We applied the QPE procedure to FHP for mass and momentum in x and y directions. For each procedure we used the same operators as previously defined , with elements $u_q(s,s') = \delta_{s,s'} e^{i q(s)}$. The results we obtained are shown in Figure \ref{fig:qpe_fhp}. 
Analogous simulations were run for momentum in x and y direction. The success of the algorithm can be qualitatively seen by the presence of one peak for each mass.
We proved that a phase estimation algorithm offers an alternative to retrieving quantities of interest. This procedure can be equally applied to any \dnqv model. 

\section{\label{sec:conclusion} Conclusion}

In this paper, we propose different ideas and methodologies for executing LGCA on quantum computers. In the first place, we highlight the limitations of an advantageous quantum representation of classical \dnqv models using CBE, expanding previous results. We show that the limitation is not given specifically by the CBE adopted, but relies also on the definition of the streaming operator. This opens perspectives on future directions to consider for moving towards quantum advantage. We develop and prove the validity of different methodologies for finding quantum collision circuits. In particular, we develop a new collision circuit for D1Q3, proving the validity of a method that uses QPE, and that can be expanded to any \dnqv model. Furthermore, we develop a quantum collision circuit for 0-momentum collisions of FHP. The implementation of non-deterministic collisions can be done for algorithms with CBE presenting linear or sublinear encoding of the space, providing reinitialization in this second case. This is due to the measurement of an ancillary qubit as a quantum analogy to a classical random extraction. Our algorithm for FHP collisions proves to be optimal in the number of universal gates needed, with a lower cost, compared to the default Qiskit decomposition, of $\approx 100$ times. The method we use, consisting of relying on classical features of the verification process, can be further applied to finding optimal circuits for executing other collisions. We develop and prove the effectiveness of a method for counting quantum invariants, showing unexpected results that contribute to the study of quantum \dnqv models. This method can be applied to any unitary collision operator that can be defined for any \dnqv model. A method for quantum invariants of non-unitary operators is left as a future perspective. Finding an explanation and the possible consequences of the presence of numerous quantum spurious invariants will be subject to further studies. In the last place, QPE was used for the first time as a subroutine for detecting quantities of interest using CBE. This procedure can be carried out in quantum algorithms that use CBE, with any encoding of the space, and for any purpose that involves moving physical information to an additional register. All the methods here presented can be applied to quantum algorithms that involve CBE, or that generally use an orthogonal set of quantum states for representing different classical occupation states.

\section*{Acknowledgments}
This work is supported by the PEPR integrated project EPiQ ANR-22-PETQ-0007, by the ANR JCJC DisQC ANR-22-CE47-0002-01 founded from the French National Research Agency and with the support of the french government under the France 2030 investment plan, as part of the Initiative d'Excellence d'Aix-Marseille Université - A*MIDEX AMX-21-RID-011.

\section*{Authors' contribution}
\textbf{Niccolo Fonio}: conceptualization, data curation, methodology, formal analysis, investigation. \textbf{Giuseppe di Molfetta} and \textbf{Pierre Sagaut}: conceptualization, funding acquisition, writing - review \& editing. All authors contributed to write, read and approved the final manuscript.

\section*{Declaration of competing interest}
The author declare no competing interests

%%      ~\cite{<label>} ==> [1]

%%

%% \bibliographystyle{elsarticle-num} 
%% \bibliography{bibliography}



\appendix

\section{Sublinear encoding of the space}
\label{app2}

Observing the possibilities of a sublinear encoding of the space is of major interest for reducing drastically the number of qubits required, giving a practical advantage for these algorithms since they rely on large grids. Previous works~\cite{budinski2021quantum, budinski2021quantum2} have provided quantum algorithms for LBM simulations. Their remark is to practice the streaming step with a quantum walk procedure. We show that the same procedure applied to LGCA results in an exact evolution of the system, but constrains the collision step, as partly treated in~\cite{schalkers2024importance}. We show, expanding this previous result, that a unitary streaming is possible does not allow a collision step for the intrinsic indistinguishability of the cell imposed. 

The most general sublinear encoding of the lattice is

\begin{equation} \label{eq:lattice_t}
    \ket{\Psi(t)} = \frac{1}{\sqrt{N}} \sum_x \ket{x} \ket{\psi(x,t)}
\end{equation}
where $N$ is the number of cells.
The information about the cell is contained in $\ket{\psi(x)}$. We can see that each cell needs to have some fundamental features. There must be information about each velocity and information about the occupation state of the corresponding velocity. The separability of velocities allows us to use a quantum walk procedure. The separability of the occupation state is necessary for the execution of arbitrary collisions. One example is the following encoding of the cell. The first register $\ket{v}$ represents the velocity, while the second qubit represents the occupation state being $V$ the number of velocities, 

\begin{equation} \label{eq:encod_vn}
    \ket{\psi(x,t)} = \frac{1}{\sqrt{V}} \sum_v \ket{v}\ket{n_v(x,t)}
\end{equation}

With this encoding, we have a correspondence between any classical state of the cell and the quantum counterpart, as we can see in this example of D1Q2 model, equal to D1Q3 without rest particle.
\begin{align}
    [00] & \longrightarrow \ket{\psi_{00}} = \frac{\ket{00}+\ket{10}}{\sqrt{2}} = \ket{+0} \label{00} \\
    [01] & \longrightarrow \ket{\psi_{01}} = \frac{\ket{00}+\ket{11}}{\sqrt{2}} = \ket{\beta_{00}} \label{01}\\
    [10] & \longrightarrow \ket{\psi_{10}} = \frac{\ket{01}+\ket{10}}{\sqrt{2}} = \ket{\beta_{01}} \label{10}\\
    [11] & \longrightarrow \ket{\psi_{11}} = \frac{\ket{01}+\ket{11}}{\sqrt{2}} = \ket{+1} \label{11}
\end{align}

The direct advantage of an encoding like this is the streaming step, which can be translated in a series of controlled shift operators, implementing a quantum walk with the following operator

\begin{equation} \label{eq:streaming}
    \hat{S} = \sum_v \hat{\Delta}_v \otimes \ket{v}\bra{v} \otimes I = \sum_v \hat{\Delta}_v \otimes \hat{I}_v
\end{equation}
Where $\hat{I}_v=\ket{v}\bra{v}\otimes \hat{I}$, and the identity acts on the occupation register because the streaming does not change the information about the occupation register, but only its location. Also,

$$
\hat{\Delta}_v =\sum_x \ket{x+v}\bra{x}
$$

The streaming procedure does not apply only to the encoding encoding in Eq. \ref{00}-\ref{11}, which is not considered in the following discussion. If we apply $S$ to the lattice, in the cell-register we are going to have the subsystem of the cell that is divided depending on the velocity, according to the streaming operator. 

\begin{align*}
    \ket{\Psi(t+1)} & = \hat{S}\ket{\Psi(t)} \\
    & = (\sum_v \hat{\Delta}_v \otimes \hat{I}_v) \frac{1}{\sqrt{N}} \sum_x \ket{x} \ket{\psi(x,t)} \\
    & = \frac{1}{\sqrt{N}} \sum_x  \sum_v \hat{\Delta}_v \ket{x} \otimes \hat{I}_v \ket{\psi(x,t)} \\
    & = \frac{1}{\sqrt{N}}  \sum_x    \sum_v\ket{x+v} \otimes \hat{I}_v \ket{\psi(x,t)}
\end{align*}

In general, we can write
\begin{align*}
    \ket{\Psi(t+1)} & = \sum_x \ket{x} \otimes \sum_v \hat{I}_v\ket{\psi(x-v,t)} \\
    & = \sum_x \ket{x} \otimes \ket{\psi(x,t+1)}
\end{align*}
Where 
\begin{equation} \label{eq:post_streaming_2}
    \ket{\psi(x,t+1)} = \sum_v \hat{I}_v\ket{\psi(x-v,t)}    
\end{equation}

The Eq.\ref{eq:post_streaming_2} depends directly on the choice of performing a quantum walk for moving the information through the lattice. It says that the information after the streaming (lhs) merges information before the streaming (rhs). We can show that this condition, which holds for any \dnqv model with the general sublinear encoding presented, forbids the post-streaming state to belong to an orthogonal set, thus forbidding a unitary collision. To prove this, we consider the D1Q2 case, but the same procedure can be extended to other \dnqv models

We define the most general encoding as follows, in matrix notation for simplicity, we consider a cell register of 2 qubits

\begin{equation} \label{eq:gen_enc}
    \ket{\psi_{i,j}} = 
    \begin{bmatrix}
    a_{i,j} \\
    b_{i,j} \\
    c_{i,j} \\
    d_{i,j} 
\end{bmatrix}
\end{equation}

Using $\ket{v_0} = \ket{0}, \ket{v_1} = \ket{1}$ and we try to solve Eq.\ref{eq:post_streaming_2}.
We can formulate Eq.\ref{eq:post_streaming_2} for D1Q2 as follows $\forall k,k' \in {0,1,2,3}$

\begin{equation} \label{eq:post_stream_2}
    \ket{\psi_{ij}} = \hat{I}_0 \ket{\psi_{ik}} + \hat{I}_1 \ket{\psi_{k'j}}
\end{equation}

This is a set of 16 equations. Some of them are trivial and if we  use Eq.\ref{eq:gen_enc}, we find that there are 8 conditions to satisfy
\begin{align}
    c_{00} = c_{10} & & d_{00} = d_{10} & & a_{00} = a_{01} & & b_{00} = b_{01} \notag \\
    c_{11} = c_{01} & & d_{11} = d_{01} & & a_{11} = a_{10} & & b_{11} = b_{10} \label{eq:post streaming condition}
\end{align}

We can rewrite our states as follows

\begin{align*}
    \ket{\psi_{00}} = 
    \begin{pmatrix}
    a \\
    b \\
    c \\
    d
    \end{pmatrix} & & 
    \ket{\psi_{01}} = 
    \begin{pmatrix}
    a \\
    b \\
    e \\
    f
    \end{pmatrix} \notag \\
    \ket{\psi_{10}} = 
    \begin{pmatrix}
    g \\
    h \\
    c \\
    d
    \end{pmatrix} & &
    \ket{\psi_{11}} = 
    \begin{pmatrix}
    g \\
    h \\
    e \\
    f
    \end{pmatrix}
\end{align*}

Now we can apply the orthogonality condition and we get these 6 equations 
\begin{equation} \label{eq:orthogonality}
    \bra{\psi_{i,j}}\ket{\psi_{i',j'}}=\delta_{i,i'}\delta_{j,j'}    
\end{equation}
We apply the following definitions to rewrite \ref{eq:orthogonality} and normalization conditions
\begin{align*}
    A & = |a|^2 + |b|^2 \\
    B & = |c|^2 + |d|^2 \\
    C & = |e|^2 + |f|^2 \\
    D & = |g|^2 + |h|^2 \\
    E & = a^* g + b^* h \\
    F & = c^* e + d^* f
\end{align*}
And we get the following set of equations
\begin{align*}
    && A + F = 0 && E + B = 0 && E + F = 0 \\ 
    && E + F^* = 0 && E + C = 0 && D + F = 0 \\
    A + B = 1 && A + C = 1 && B + D = 1 && C + D = 1
\end{align*}
The first two lines are the orthogonality conditions, and the last line are the normalization conditions. It is easy now to see that these equations are not solvable, since according to the first three $A+B=0$, while the normalization condition imposes $A+B=1$. The peculiar aspect is that even increasing the space, i.e. increasing the qubits for an encoding, brings to the same set of equations. This is intrinsically caused by the streaming operator that separates the information, making it impossible to merge it in a distinguishable way. These considerations are intended as a hint that a sublinear encoding preserving orthogonality in the cell register and performing the evolution of LGCA through quantum walk is not possible.

In conclusion, this encoding does not permit the practice of the collision step, for the impossibility of ensuring the orthogonality of the cell states. Our results supposed for simplicity to have the case of the chosen velocities $\ket{0}$ and $\ket{1}$, and of the streaming operator Eq.\ref{eq:streaming}. The same proof holds in case of different velocities, that must anyway always be orthogonal. This orthogonality of the velocities, needed for the streaming procedure, brings to similar equations to Eqs\ref{eq:post streaming condition} for any DnQv model. A general proof considering an alternative and more general streaming is yet to be found. Alternative streaming would change the character of the algorithm, which is conceived as a naive translation of the classical algorithm using a superposition.

The circuit that performs the streaming procedure we just explained, considering the example encoding in Eq. \ref{00}-\ref{11}, is given in Fig.\ref{fig:qcircuit_d1q2}. Here the register $\ket{x}$ is the space register. The cell occupational state is represented with two qubits $\ket{v}$ and $\ket{n}$. $\ket{v}$ has the information about the velocity ($\ket{0} \rightarrow v=+1$ and $\ket{1} \rightarrow v=-1$) and will be used by the streaming operator. $\ket{n}$ has information about the occupational state of the corresponding velocity. The classical-quantum correspondence is in Eq. \ref{00}-\ref{11}. First, the space register undergoes a series of Hadamard gates, creating the superposition. Then the operation $A_0$ is the initialization procedure, which is a series of generalized Toffoli gates with space register qubits and $\ket{v}$ as controls, and $\ket{n}$ as target. More optimal operations could be found \cite{shende2005synthesis}. Then, the operators $\Delta_r$ and $\Delta_l$ changes the position according to the velocity state of $\ket{v}$, which is the control. This collisionless multi-time step algorithm was run for 24 time steps and the results are shown in Fig.\ref{fig:qd1q2_sim}.

\begin{figure}[ht]
\[
\begin{array}{c}

\Qcircuit @C=1.0em @R=0.2em @!R {
    \lstick{\ket{x_0}} & \gate{H} & \multigate{5}{A_0} & \multigate{3}{\Delta_r} & \multigate{3}{\Delta_l} & \qw \\
    \lstick{\ket{x_1}} & \gate{H} & \ghost{A_0} & \ghost{\Delta_r} & \ghost{\Delta_l} & \qw\\
    \lstick{\ket{x_2}} & \gate{H} & \ghost{A_0} & \ghost{\Delta_r} & \ghost{\Delta_l} & \qw\\
    \lstick{\ket{x_3}} & \gate{H} & \ghost{A_0} & \ghost{\Delta_r} & \ghost{\Delta_l} & \qw \\
    \lstick{\ket{v}} & \qw & \ghost{A_0} & \ctrlo{-1} & \ctrl{-1} & \qw\\
   \lstick{\ket{n}} & \qw & \ghost{A_0} & \qw & \qw & \qw
}
\end{array}
\]
\caption{Quantum circuit with unitary steaming for D1Q2. The $x$ register is the space register, $v$ is the velocity register and $n$ is the occupation register. Here we count $2^4=16$ sites. The $A_0$ operation is a series of generalized Toffoli that initialize the occupation qubits in the $n$ register. The controlled $\Delta_v$ operations are repeated for each time step} \label{fig:qcircuit_d1q2}
\end{figure}

\begin{figure}[ht]
    \centering
    \includegraphics[scale=0.25]{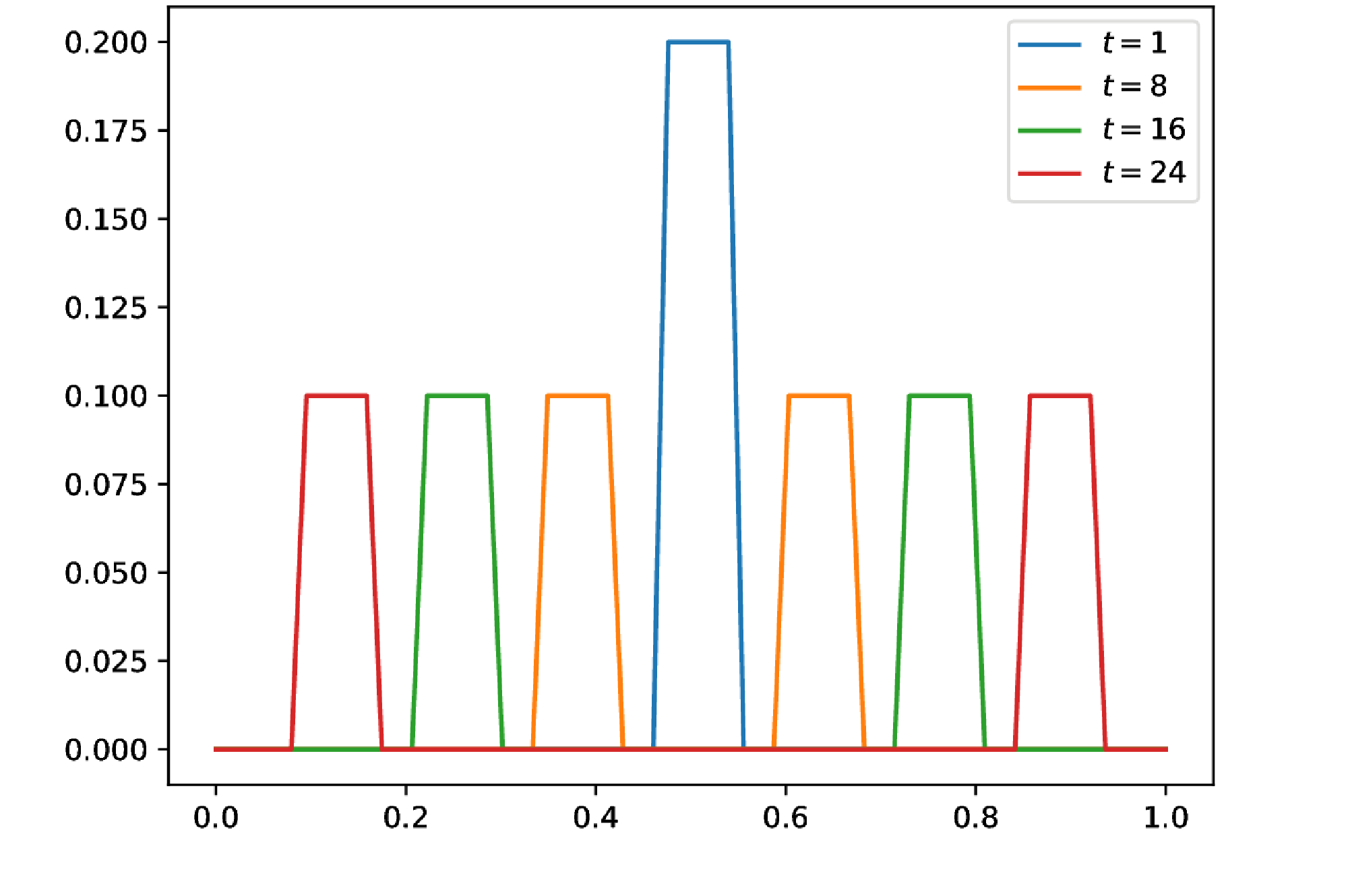}
    \caption{D1Q2 lattice mass density at different time-steps with 64 cells run with the quantum circuit in Fig.\ref{fig:qcircuit_d1q2}. This simulation was run with mps simulator by Qiskit. At increasing time steps we can see the system diffuses and particles move according to respective velocities. After each value of $t$ the system was measured with 1000 shots.}
    \label{fig:qd1q2_sim}
\end{figure}

\section{Complete evolution of D1Q3 operators}

In Table\ref{tab:d1q3_evolution} we give all the evolved operators as calculated with the collision provided by Love. Thus, it is possible to verify directly the conservation of the quantities reported in the article. Analogous evolved operator for FHP can be provided upon request

\begin{table}[ht]
\resizebox{\textwidth}{!}{
\begin{tabular}{cl|cl}
\textrm{Input state} $\hat{O}$&
\textrm{Output state} $\hat{C}^{\dagger}\hat{O}\hat{C}$&
\textrm{Input state} $\hat{O}$&
\textrm{Output state} $\hat{C}^{\dagger}\hat{O}\hat{C}$\\
III & III & XZZ & $\frac{1}{2}$(-IXX - IYY + XII + XZZ) \\
IIX & $\frac{1}{2}$(IIX + XXI + YYI + ZZX) & YII & $\frac{1}{2}$(-IXY + IYX + YII + YZZ) \\
IIY & $\frac{1}{2}$(IIY + XYI - YXI + ZZY) & YIX & $\frac{1}{2}$(IYI - XIY + YIX + ZYZ) \\
IIZ & $\frac{1}{2}$(IIZ + IZI - ZII + ZZZ) & YIY & $\frac{1}{2}$(-IXI + XIX + YIY - ZXZ) \\
IXI & $\frac{1}{2}$(IXI + XIX - YIY - ZXZ) & YIZ & $\frac{1}{2}$(YIZ + YZI + ZXY - ZYX) \\
IXX & $\frac{1}{2}$(IXX - IYY + XII - XZZ) & YXI & $\frac{1}{2}$(-IIY + XYI + YXI + ZZY) \\
IXY & $\frac{1}{2}$(IXY + IYX - YII + YZZ) & YXX & $\frac{1}{2}$(-XXY + XYX + YXX - YYY) \\
IXZ & $\frac{1}{2}$(IXZ + XZX - YZY - ZXI) & YXY & YXY \\
IYI & $\frac{1}{2}$(IYI + XIY + YIX - ZYZ) & YXZ & $\frac{1}{2}$(-IZY + XYZ + YXZ + ZIY) \\
IYX & $\frac{1}{2}$(IXY + IYX + YII - YZZ) & YYI & $\frac{1}{2}$(IIX - XXI + YYI - ZZX) \\
IYY & $\frac{1}{2}$(-IXX + IYY + XII - XZZ) & YYX & YYX \\
IYZ & $\frac{1}{2}$(IYZ + XZY + YZX - ZYI) & YYY & $\frac{1}{2}$(-XXY + XYX - YXX + YYY) \\
IZI & $\frac{1}{2}$(IIZ + IZI + ZII - ZZZ) & YYZ & $\frac{1}{2}$(IZX - XXZ + YYZ - ZIX) \\
IZX & $\frac{1}{2}$(IZX + XXZ + YYZ + ZIX) & YZI & $\frac{1}{2}$(YIZ + YZI - ZXY + ZYX) \\
IZY & $\frac{1}{2}$(IZY + XYZ - YXZ + ZIY) & YZX & $\frac{1}{2}$(IYZ - XZY + YZX + ZYI) \\
IZZ & IZZ & YZY & $\frac{1}{2}$(-IXZ + XZX + YZY - ZXI) \\
XII & $\frac{1}{2}$(IXX + IYY + XII + XZZ) & YZZ & $\frac{1}{2}$(IXY - IYX + YII + YZZ) \\
XIX & $\frac{1}{2}$(IXI + XIX + YIY + ZXZ) & ZII & $\frac{1}{2}$(-IIZ + IZI + ZII + ZZZ) \\
XIY & $\frac{1}{2}$(IYI + XIY - YIX + ZYZ) & ZIX & $\frac{1}{2}$(IZX - XXZ - YYZ + ZIX) \\
XIZ & $\frac{1}{2}$(XIZ + XZI - ZXX - ZYY) & ZIY & $\frac{1}{2}$(IZY - XYZ + YXZ + ZIY) \\
XXI & $\frac{1}{2}$(IIX + XXI - YYI - ZZX) & ZIZ & ZIZ \\
XXX & XXX & ZXI & $\frac{1}{2}$(-IXZ + XZX - YZY + ZXI) \\
XXY & $\frac{1}{2}$(XXY + XYX - YXX - YYY) & ZXX & $\frac{1}{2}$(-XIZ + XZI + ZXX - ZYY) \\
XXZ & $\frac{1}{2}$(IZX + XXZ - YYZ - ZIX) & ZXY & $\frac{1}{2}$(YIZ - YZI + ZXY + ZYX) \\
XYI & $\frac{1}{2}$(IIY + XYI + YXI - ZZY) & ZXZ & $\frac{1}{2}$(-IXI + XIX - YIY + ZXZ) \\
XYX & $\frac{1}{2}$(XXY + XYX + YXX + YYY) & ZYI & $\frac{1}{2}$(-IYZ + XZY + YZX + ZYI) \\
XYY & XYY & ZYX & $\frac{1}{2}$(-YIZ + YZI + ZXY + ZYX) \\
XYZ & $\frac{1}{2}$(IZY + XYZ + YXZ - ZIY) & ZYY & $\frac{1}{2}$(-XIZ + XZI - ZXX + ZYY) \\
XZI & $\frac{1}{2}$(XIZ + XZI + ZXX + ZYY) & ZYZ & $\frac{1}{2}$(-IYI + XIY + YIX + ZYZ) \\
XZX & $\frac{1}{2}$(IXZ + XZX + YZY + ZXI) & ZZI & ZZI \\
XZY & $\frac{1}{2}$(IYZ + XZY - YZX + ZYI) & ZZX & $\frac{1}{2}$(IIX - XXI - YYI + ZZX) \\
XZZ & $\frac{1}{2}$(-IXX - IYY + XII + XZZ) & ZZY & $\frac{1}{2}$(IIY - XYI + YXI + ZZY) 
\end{tabular}
}
\caption{\label{tab:d1q3_evolution}
Table of the evolution of all the D1Q3 Pauli operators
}
\end{table}

\end{document}